\documentclass[12pt]{iopart}

\usepackage[hyphens]{url}
\usepackage[top=2cm,bottom=3cm,left=2.4cm,right=2.4cm]{geometry}
\usepackage[usenames,dvipsnames,table]{xcolor}
\usepackage[titletoc]{appendix}
\usepackage[makeroom]{cancel}
\usepackage{graphicx}
\usepackage[color,final]{showkeys}
\usepackage{epstopdf}
\usepackage{mathrsfs}
\usepackage{enumitem}
\usepackage{stackrel}
\usepackage{float}
\usepackage{booktabs}
\usepackage{xcolor}
\usepackage[normalem]{ulem}  
\usepackage{hyperref}
\usepackage{longtable}
\usepackage{cite} 
\usepackage{comment}
\usepackage{blkarray}
\usepackage{multirow}
\usepackage{footmisc}

\begin{document}

\def\ben{\begin{equation}}
\def\een{\end{equation}}
\def\be{\begin{equation}}
\def\ee{\end{equation}}
\def\bea{\begin{eqnarray}}
\def\eea{\end{eqnarray}}
\def\bA{{\bar{A}}}
\def\vx{{\vec{x}}}
\def\vy{{\vec{y}}}
\def\br{{\bar{r}}}
\def\hO{{\hat{O}}}
\def\cO{{\cal{O}}}
\def\tX{{\tilde{X}}}
\def\tP{{\tilde{P}}}
\def\fr{\frac}
\def\lan{\langle}
\def\ran{\rangle}
\def\mH{{\mathcal H}}
\def\bA{{\bar A}}
\def\eq#1{(\ref{#1})}
\def\l{\lambda}
\def\gap#1{\vspace{#1ex}}
\def\vev#1{\langle{#1}\rangle}

\def\red#1{{\color{red}{\bf #1}}}

\begin{flushright}
	TIFR/TH/20-8
\end{flushright}

\title[Bulk Entanglement Entropy and Matrices]{\kern70pt Bulk Entanglement Entropy and Matrices}

\author{Sumit R. Das $^1$, Anurag Kaushal $^2$, Gautam Mandal $^2$ and  \\\ Sandip P. Trivedi $^2$}
\address{$^1$ Department of Physics and Astronomy, University of Kentucky, Lexington, KY 40506, USA. \\
$^2$ Department of Theoretical Physics, Tata Institute of Fundamental Research, Mumbai 400005, INDIA.}
\ead{das@pa.uky.edu, anurag.kaushal@theory.tifr.res.in, mandal@theory.tifr.res.in, sandip@theory.tifr.res.in}
\vspace{10pt}
\begin{indented}
\item[]
\end{indented}

\gap5

\begin{center}
  {\it In memory of Peter Freund}\\
\end{center}

\gap5

\begin{abstract}
Motivated by the Bekenstein Hawking formula and the area law behaviour of entanglement entropy, we propose  that in any  UV finite theory of quantum gravity with  a smooth spacetime, the total entropy  for  a pure state in a co-dimension one spatial region, to leading order,  is given by $S={A\over 4 G_N}$, where $A$ is the area of the co-dimension two boundary. 
 In the context of $Dp$ brane holography we show that  for some specially chosen regions  bulk  entanglement  can be mapped to ``target space" entanglement in the boundary theory. Our conjecture  then  leads to a precise proposal  for target space entanglement in the boundary theory at strong coupling and large $N$. In particular it  leads to the conclusion that the target space entanglement would scale like $O(N^2)$ which is quite plausible in a system with $O(N^2)$ degrees of freedom. Recent numerical advances in studying the D0 brane system hold out the hope that this proposal can be tested in a precise way in the future.


\end{abstract}

%
%
%
%
%

\newpage

\section{Introduction}

Quantum entanglement plays a key role in gauge-gravity duality. In AdS/CFT correspondence, the Ryu-Takayanagi formula \cite{Ryu:2006bv} and its covariant version \cite{Hubeny:2007xt}, together with its extensions \cite{Faulkner:2013ana,Engelhardt:2014gca}, provide a strikingly simple geometric understanding of the entanglement entropy of a subregion in the boundary theory in terms of extremal surfaces in the bulk. 

In this note we consider entanglement entropy in the bulk itself
and its interpretation in the boundary theory. Consider some spatial subregion of the bulk and the entanglement of this subregion with its complement. We may consider the associated entanglement entropy. Our aim is to ask:  what is the value for this ``bulk entanglement entropy" and what is the meaning of this quantity in the dual quantum field theory? This bulk entanglement entropy is itself, of course, a tricky concept since the bulk theory is a theory of gravity. Nevertheless one would expect that in some smooth spacetime background with approximately local physics, this is the entanglement of quantum fields, including gravitons, across the co-dimension two boundary of the subregion. Our aim is to seek a definition of this quantity which agrees with this notion in the semiclassical regime, but can be extended beyond this. We will propose such a definition in terms of the holographically dual field theory.

The leading term in this entropy is expected to be proportional to the area of the boundary in units of an appropriate power of a UV cutoff. In the context of entanglement across a black hole horizon, it has been argued that this UV divergent term renormalizes the Newton constant \cite{thooft,susskind1,susskind2}.  In an UV complete theory of gravity the answer should be finite, and it is natural to then ask what provides this cutoff. 

In this note we conjecture that given a  consistent theory of quantum gravity, in any smooth spacetime the entanglement entropy of a spatial co-dimension one region is given by the area formula, 
\ben
\label{seere}
S_{EE}={A_E\over 4 G_N},
\een
where  $A_E$ denotes the area in the Einstein frame. Eq. \eq{seere} is meant to provide the leading order behaviour of the entanglement entropy, with possible subleading corrections, e.g.,  arising in string theory from $g_s$ and $\alpha'$ corrections, which may not be universal. 

In the above we assumed a pure state for the whole system. In case of a mixed state, the total von Neumann entropy of the region, has, in addition to the entanglement entropy, an extra, {\it classical} or {\it disorder}, contribution, as explained shortly. We believe that, for mixed states, the formula above continues to hold for the {\it total} entropy 
\ben
\label{conj}
S={A_E\over 4 G_N}.
\een
We expect \eq{conj} to be true where the density matrix for the mixed state for the full system can be obtained starting from a pure state in a bigger system which also admits a gravity description (indeed in such cases (\ref{conj}) follows from the conjecture for pure states, (\ref{seere})). Such a situation arises for a thermal state in AdS/CFT which is described by a black hole in the gravity dual. The thermal state's density matrix can be obtained from that of a pure state in the thermo-field double system and the corresponding dual geometry is the double sided, eternal, black hole.
More generally for mixed states we suspect that some version of eq.(\ref{conj}) is still valid, once it is  stated precisely in terms of the fine-grained entropy of the degrees of freedom of  a light sheet associated with a boundary of area $A$ \cite{Bousso},  but we have not understood this  well yet and leave a detailed discussion for   general mixed states for the future \footnote{We are grateful to R. Bousso for drawing our attention to this issue and in particular for bring reference  \cite{Bousso} to our notice.}. 

We can be most definite about our conjecture eq.(\ref{seere}) in the context of AdS/CFT like situations where the gravity theory has a dual which lives on a time- like boundary. 
In such situations, which arise for example in the near-horizon geometries of $Dp$ brane theories, the state in the bulk can be mapped to a state in the boundary dual. 
The bulk entanglement then  maps to the von Neumman entropy of a suitably defined density matrix in the boundary theory, which we will see shortly is associated with target space entanglement. Even in such situations though,  once a black hole is present, one can consider a bulk region inside the horizon and the precise map of the bulk entanglement to  the boundary theory is such cases is not well understood by us  as yet and left for further investigation. 



Our conjecture stems from the following, admittedly intuitive, reasoning. One expects in any quantum mechanically complete theory to get a finite result for the bulk entanglement, for example, in a  closed string theory describing the bulk. For a black hole horizon one also expects the result to agree with the Bekenstein-Hawking formula,
\ben
\label{sbh}
S_{BH}={A_H\over 4 G_N}.
\een 
This suggests that the cut-off rendering the entanglement finite is provided by $G_N$ in general, leading to the relations  eq.(\ref{conj}), eq.(\ref{seere}).
 In particular, for a black hole that forms from the collapse of a pure state it is quite plausible   that $S_{BH}$  is accounted for completely by  entanglement, leading to   eq.(\ref{seere}).

Let us note that in the context of Einstein gravity with some matter, the entanglement entropy of matter and gravitons across the black hole horizon appears as a quantum correction to the Bekenstein-Hawking entropy which may be regarded as a "classical" contribution \cite{susskind2}. However if Einstein gravity itself is an effective theory obtained by integrating out massive closed string modes, such a classical contribution itself can be considered as an entanglement entropy of the fundamental degrees of freedom. This viewpoint is consistent with what happens in models of induced gravity \cite{induced}.

We also note that our conjecture eq.(\ref{conj}) is equivalent to saying  that the Bekenstein bound is  saturated by our notion of the bulk entropy,  to leading order, in any smooth background. For mixed states, as indicated above \eq{conj}, the total entropy is a sum of ``quantum'' entanglements and a ``disorder'' part inherited from the mixed state. Recall that for a quantum system, an initial mixed state $\rho$ $=\sum_i w_i |\alpha_i\ran \lan \alpha_i|$,  the form of the reduced density matrix is $\rho_A = \sum_i w_i \rho_A^{i}$ whose von Neumann entropy combines the quantum entanglement from the $\rho_A^{i}= \tr_{A^c}|\alpha_i\ran \lan \alpha_i|$ (inherent in the pure states $|\alpha_i\ran$) with the classical or  ``disorder'' contribution $-\sum_i w_i \log w_i$ which would be present even if the states $|\alpha_i\ran$ are factorizable states. For target space entropies, which will be of relevance for us, these concepts are suitably generalized. We envisage the saturation of the bound, if (\ref{conj}) holds for mixed states, as a  trade off between these two parts of the entropy such that the sum equals the bound.

What makes the conjecture above interesting is that improvements in numerical techniques now hold out the hope that we can   test it precisely in the future. 
With this motivation in mind  here we consider this problem in the context of $Dp$ brane holography for some special spatial co-dimension one regions. We show  that the bulk entanglement entropy can be mapped to the boundary theory in a fairly precise manner. The boundary theory, for example  for $D0$ branes, and more generally  for $Dp$ branes with $p<3$ , has no dimensionless parameter other than $N$ - the number of branes. This constrains the form of the result,  and one finds that the expression in eq.(\ref{seere}) agrees with what  
could arise in the boundary theory. In  fact the result eq.(\ref{seere}),  when expressed in terms of the appropriate dimensionless variables, scales like $N^2$ which is quite plausible in a system with $O(N^2)$ degrees of freedom. 





We find that for the special co-dimension one regions we consider the bulk entanglement  maps in  a fairly precise way to a quantity sometimes referred to  as ``target space entanglement" in the boundary theory. It is worth pausing to briefly  explain this idea here. 
Consider a quantum mechanical system where the degrees of freedom live in time alone. Some of these degrees of freedom include target space directions along which the system can move, such a system arises for example in the case of the field theory limit of the $D0$ brane theory. There is no spatial extent in the quantum mechanical case so we cannot consider a spatial sub -region and define an entanglement in that manner, as is often done in a field theory. However  we can consider some restriction  in the target space and associate an entanglement  with this restriction -this  is  referred to as the target space entanglement. The simplest example is  a single particle, say an harmonic oscillator, in one dimension $x$; we may want to restrict ourselves to some region $a<x<b$ and only concern ourselves with measurements which can be made when we restrict ourselves to this region. Even if the state of the system is pure, this restriction on the set of all observables we have access to gives rise to a density metric whose von Neumann entropy is then the  target space entanglement entropy. If $\Psi(x)$ is the wave function, the density matrix is given in the position basis by 
\ben
\label{defrho}
\rho(x,x')=\Psi^*(x) \Psi(x')
\een
with $x\in[a,b]$.
The sub algebra of observables one is correspondingly restricted to is given by operators of the form
\ben
\label{defog}
{\hat O}=\int_{a}^{b} \int_a^b dx dx'C(x,x')|x\ran\lan x'|
\ee where $C$ is hermitian, satisfying, $C(x,x')=C^*(x',x)$. One can think of the entanglement entropy as being associated with this sub algebra.
While target space entanglement agrees with the usual notion of entanglement of a bulk region in the regime of couplings where the bulk is semi-classical, it clearly remains a well-defined quantity for any regime of couplings. The target space entanglement, therefore, provides the general notion of "bulk entanglement" that  we are seeking.

Target space entanglement has been implicitly used to define notions of entanglement entropy in several situations. One example involves worldsheet formulations of string theory \cite{susskind1,stringee}. It is also the basis for discussion of entanglement entropy in the $c=1$ matrix model \cite{cone} dual to two dimensional non-critical string theory \cite{Das:1995vj, Hartnoll:2015fca} This also appears in a slightly different context in \cite{internalextremal}. The formalism of general target space entanglement has been recently developed in \cite{DMT:2018}, \cite{Mazenc:2019ety}.

In general due to the non-commuting nature of the target space spatial coordinates in the $D0$ brane theory (and similarly the  higher $Dp$ systems) it is not possible to precisely 
map a region in the bulk to an appropriate restriction in the target space of the boundary theory. However for some  carefully chosen co dimension one regions in the bulk we show that this is possible. This then allows us to map the bulk entanglement entropy quite precisely to target space entropy in the boundary theory. 

Our mapping is not totally precise though, and we find that  there are two  natural possibilities which  arise in the boundary theory. Distinguishing between them and checking whether the target space entanglement in either case  agrees with (\ref{seere})  would require numerical work. In fact there have been great strides recently in studying some of the field theories which arise in the context of $AdS/CFT$ numerically. For example the free energy at strong coupling for the $D0$ brane matrix theory has been studied by \cite{Hanada:2016zxj} and shown to agree quite precisely with the bulk result coming from a black hole. 
While calculations of target space entanglement will be much more challenging, these advances allow us to hope that such a calculation can be carried out in the not so distant future, allowing a test of  whether either of the two possibilities for the target space entanglement agrees with eq.(\ref{seere}). Such a numerical calculation would provide  a very non-trivial check for our conjecture
(For a recent discussion of entanglement entropy in matrix models in a different context see \cite{alet}).


Some of the above discussion is best understood for the $c=1$ model which is dual to $1+1$ dimensional string theory \cite{cone}. Here the space in the string theory arises from the space of eigenvalues of the $N \times N$ hermitian matrix $M$ while the eigenvalues themselves are coordinates of $N$ fermions. The "bulk" description arises from second quantization of these fermions. The fermion field can be bosonized yielding collective field theory of the density of eigenvalues. The fluctuations of the collective field are related to the "massless tachyon" of the two dimensional string theory, which is the only dynamical mode. The bulk entanglement entropy of an interval has the usual meaning in this second quantized language and was computed in \cite{Das:1995vj}. This calculation has been more recently revisited and improved in \cite{Hartnoll:2015fca}. The leading answer for the entanglement entropy of an interval is {\em finite}; the UV cutoff discussed earlier is provided here by the position dependent string coupling. The fact that the string coupling enters as the cutoff is consistent with the conjecture that the Newton constant provides the UV cutoff. The finiteness can be ultimately traced to the fact that we are dealing with $N \times N$ matrices. In the bosonic formulation, this manifests itself in the fact that there are at most $N$ independent single trace operators of the form ${\rm Tr}M^n$. (Note that ${\rm Tr}M^n$ for $n>N$ is expressible as a sum of products of the lower single traces.) Since $n$ is a quantum number conjugate to the emergent space direction, this means that the collective field should really be thought of as living on a lattice with spacing $\sim 1/N$. This becomes clearer in a basis formed by the characters of the permutation group which are in one to one correspondence to fermion wavefunctions \cite{Jevicki:1991yi}, or in the formalism of bosonization of a finite number of fermions in \cite{dhar}. Matrix quantum mechanics is equivalent to the first quantized formulation - the bulk entanglement then relates to an appropriate subalgebra of operators. This becomes an example of "target space entanglement"  mentioned above.

Our considerations also apply to field theories, which arise for example as duals in the $Dp$ brane case with $p \le 3$. 
In this case there is the  the usual notion of entanglement entropy associated with a spatially localized region. However one can also consider a  notion of target space entropy which  arises when one restricts to observables which can only access some region of target space, without imposing any restriction along the spatial directions in which the field theory lives. We show that it is the latter type of target space entanglement which is dual to the bulk entanglement when we consider spatial regions in the bulk extending fully along those in which the field theory lives with restrictions only in the spatial directions transverse to the field theory ones. 
Upto the kind of ambiguity mentioned above which one faces in the D0 brane case, the mapping of the bulk entanglement to the field theory target space entanglement is precise, 
and we find once again that  our proposed bulk entanglement entropy eq.(\ref{seere}) scales like $N^2$ when exposed in terms of the dimensionless variables of the field theory. It is worth pointing out that  for field theories 
 we can consider  a  more general notion of entanglement where we impose a restriction on both the spatial region and within that region a further restriction on the region of target space that case be accessed. This generalized entanglement would interpolate between the usual spatially localized entanglement, which has a dual interpretation as a RT surface, and the target space entanglement we have been discussing here. We leave an exploration of this interesting idea for the future. 

As was pointed out above the formula we suggest for bulk entanglement, eq.(\ref{seere}), is only at leading order and would have corrections, due to both $\alpha'$ and string loop effects. Since the definition of target space entropy in the field theory is a general one these corrections  could  be computed on the boundary side.  We can also consider the weakly coupled limit in the field theory where the dual spacetime is highly curved with a curvature of order the string scale; the boundary theory definition would still hold in this case and would allow us to make sense of the  entanglement. 

Finally, our conjecture (\ref{seere}) considers the Einstein frame area : this is natural from the presumed connection to the Bekenstein-Hawking formula. We can also consider another possibility, viz. that the entanglement entropy is proportional to the area in the string frame metric. In fact we find that in this case we get a result which can be obtained from the holographic dual, provided we use the string length $l_s$ as the UV cutoff. However, in this case the result scales as $N^0$ rather than $N^2$. This is, in a sense, less natural to expect. However it is still possible. A detailed numerical calculation which we allude to will determine which of these alternatives is correct.

This note is organized as follows. In section 2 we calculate the bulk entanglement across a simple co-dimension two surface in the geometry of $N$ coincident D0 branes following the proposal (\ref{seere}), and show that when the parameters which appear in the setup are expressed in terms of appropriate scales of the D0 brane theory, the answer scales as $N^2$. In section 3 we put forward our proposal for the target space entanglement which corresponds to the calculation of section 2.  Section 4 extends the supergravity calculation of section 2 to Dp branes for $p < 3$. Section 5 discusses the target space entanglement proposal for the Dp brane field theory. Section 6 contains discussions of our results and their extensions. The appendices deal with the definition and evaluation of target space entanglement for the case of a single matrix relevant for $c=1$ case, and a proof that this notion is identical to the notion of entanglement in the second quantized formulation which is commonly used.

While this work was in progress, the paper \cite{vanrams} appeared, which discussed the possible relevance of areas of {\em extremal} surfaces in BFSS/gravity duality. Our work differs in an essential way : our ultimate aim to understand the meaning of entropy and entanglement across {\em any} surface in the bulk, regardless of whether it is extremal or not. (In fact in recent literature the phrase "bulk entanglement entropy" is sometimes used to talk about the entanglement across an extremal surface obtained in the context of a quantum  correction to entanglement entropy of the dual field theory associated with a spatial region of the boundary field theory \cite{Faulkner:2013ana,recentbulk}). In this paper we put forward a proposal for a set of simple surfaces in various geometries. The fact that the cutoff in the entanglement entropy in a theory of gravity should be the Newton constant has been previously argued in \cite{jacobson} from several viewpoints, in particular from the point of view of a derivation of Einsten equation as an equation of state. Our proposal is based on a consistency with a holographic description \footnote{We thank T. Jacobson for bringing this to our attention.}.
We also note that \cite{myers} had in fact proposed that the entanglement entropy in a theory of gravity saturates the Beckenstein bound  and gave some supporting evidence for the proposal which are different from ours \footnote{We thank Gary Horowitz for bringing this reference to our attention after the ArXiv version appeared.}.  Our proposal goes somewhat further: we also conjecture that for general states, the Bekenstein bound is saturated by the total entropy which includes an entanglement and a disorder part.

\section{Bulk Entanglement for D0 Brane Geometries}
\label{d0sugra}

The simplest setup is the background produced by a stack of $N$ coincident D0 branes. We begin by considering the extremal limit at temperature $T=0$. The string frame metric and the dilaton in the near horizon region are given by \cite{Itzhaki:1998dd}
\bea
ds_{string}^2 & = & -H_0(r)^{-1/2} dt^2 + H_0(r)^{1/2}[dx_1^2 + \cdots + dx_9^2], \nonumber \\
e^{-2\phi} & = & H_0(r)^{-3/2}, \nonumber \\
H_0(r) & = & \frac{R^7}{r^7}, \nonumber \\
r^2 & = & x_1^2 + \cdots x_9^2.
\label{one}
\eea
Here the scale $R$ is given by
\ben
R^7 = \frac{(2\pi)^7}{7 \Omega_8} l_s^7 (g_s N).
\label{two}
\een
$l_s$ is the string length, $g_s$ is the string coupling and $\Omega_8$ is the volume of an eight dimensional unit sphere. The string frame curvature of this solution becomes large when 
\ben
\label{condo}
r = r_0 \equiv  (g_s N)^{1/3} l_s, 
\een
so that supergravity description is valid for $r \ll r_0$. However when $r = r_1 \sim (g_s N)^{1/7} l_s$ the dilaton becomes large, so that for such small $r$ the M-theory description takes over.

Consider now dividing the nine dimensional bulk into two parts by an eight dimensional plane at $x_1 = d$. We choose $d$ to be in the region of validity of IIA supergravity or M theory, i.e. $ d \ll (g_s N)^{1/3} l_s$. 
When in addition $d \gg  (g_s N)^{1/7} l_s$, the induced string frame metric on the surface $x_1 = d$ (at a given time $t$) is
\ben
ds_{induced}^2 = H_0(\br)^{1/2} [ d\rho^2 + \rho^2 d\Omega_7^2],
\label{three}
\een
where we have defined
\bea
\rho^2 & = & x_2^2 +\cdots +  x_9^2, \nonumber \\
\br^2 & = & d^2 + \rho^2.
\label{four}
\eea
The Einstein frame area of this eight dimensional surface is then given by
\ben
A_d (T=0)= \Omega_7 \int_0^{\rho_0}d\rho\ \rho^7 H_0(\br)^{1/2} = \Omega_7 R^{7/2} \int_0^{\rho_0}d\rho\ \frac{\rho^7}{(d^2 + \rho^2)^{7/4}},
\label{five}
\een
where we have used the following relation
\[
ds_{Einstein}^2 = e^{-\phi/2} ds_{string}^2.
\]
We have imposed an IR cutoff on the integral at some $\rho_0$. We have in mind taking  
\ben
\label{condo2}
 d\ll \rho_0< r_0.
\een  
The result of the integral is then 
\ben
A_d (T=0) = \frac{2}{9} \Omega_7 R^{7/2} \rho_0^{9/2} \left[ 1 + O(d^2/\rho_0^2) \right]
\label{six}
\een
If we take the IR cutoff to be $\rho_0 \sim r_0$ and $d \sim r_1$ we see that in the regime $(g_s N) \gg 1$ the area $A_d$ behaves as (using (\ref{two}))
\ben
A_d (T=0)\sim (g_s N)^2 l_s^8 [ 1 + O((g_sN)^{-21/2})]
\label{seven}
\een

According to our proposal for the bulk entanglement, eq.(\ref{seere}) the entanglement entropy of the region $x_1 > d$ with its complement is 
\ben
S_{EE}(d) = \frac{A_d}{4 G_N}
\label{eight}
\een
As we will see soon, eq.(\ref{eight}), where the Area term is cut-off by $1/G_N$, can be expressed in terms of dimensionless quantities in the D0 brane quantum mechanics,
and will scale like $N^2$. 


Let us note in passing that one could imagine taking $d = 0$. This would necessitate including the small $r$ region of the bulk theory where the dilaton is large. This is the region described by M-theory, one expects eq.(\ref{eight}) to continue to hold in this case as well, since the RHS is invariant under a change of duality frames. 

From eq.(\ref{six}) we see that $A_d$ and therefore $S_{EE}(d)$ is 
dependent on the bulk IR cutoff $\rho_0$. We would like to get rid of this dependence so that the result can be compared in a precise way with the matrix theory. 
One way to do so  is to consider the difference between the 
entanglement entropy in a finite temperature D0 brane black hole background and the extremal D0 brane solution considered above. The near-extremal black D0 brane string frame metric is given by \cite{Itzhaki:1998dd}
\bea
ds_{string}^2 & = & -H_0(r)^{-1/2} f(r) dt^2 + H_0(r)^{1/2}[\frac{dr^2}{f(r)}+r^2 d\Omega_8^2] \nonumber \\
f(r) & = & 1- \left(\frac{r_H}{r} \right)^7
\label{ten}
\eea
while the dilaton and the one form gauge field remain the same. The horizon is now at $r=r_H$. The Hawking temperature for this solution is given by
\ben
T = \frac{7}{4\pi R} \left(\frac{r_H}{R} \right)^{5/2}
\label{eleven}
\een

Before proceeding let us note that the black hole geometry eq.(\ref{ten}) admits an extension, analogous to the well- known Kruskal extension for a Schwarzschild black hole, which has two time-like boundaries. This double sided geometry is dual to a pure state - the thermofield double state- in a system consisting of two non-interacting D0 brane systems.  
We can consider the conjecture eq.(\ref{seere})   for a bulk subregion in this extended geometry and its dual description  as the  thermo-field double state, see the comments after eq.(\ref{seere}) in the introduction.
Here we will only consider a bulk region on one side  and that too lying outside the horizon.  For our purposes therefore we do  not have to worry about the  full extended  geometry, 
and the single sided geometry,  described by the metric in eq.(\ref{ten}),  will be sufficient.
The case of  more general bulk regions is very interesting and left for the future. 
 
Consider now an $x_1=d$ surface in the geometry eq.(\ref{ten}), with 
\ben
\label{condd}
d \gg  r_H.
\een
 The area of this surface is
\ben
A_d(T) = \Omega_7 R^{7/2}\int_0^{\rho_0} d\rho~\rho^7~\frac{1}{(d^2+\rho^2)^{7/4}}~[(f(\br)^{-1}-1)\frac{\rho^2}{d^2+\rho^2}+1]^{1/2}
\label{twelve}
\een
We will consider low temperatures so that $r_H < \rho_0$. If $\rho_0 \sim r_0$ this translates to
\ben
(RT) \ll (g_sN)^{10/21}
\label{thirteen}
\een
In that case one can expand the integrand in powers of $r_H/\sqrt{d^2+\rho^2}$. To lowest order one gets
\ben
A_d(T) = \Omega_7 R^{7/2}\int_0^{\rho_0} d\rho \left[\frac{\rho^{7/2}}{(1+\frac{d^2}{\rho^2})^{7/4}} + \frac{r_H^7}{2} \frac{\rho^{-7/2}}{(1+\frac{d^2}{\rho^2})^{25/4}} + \cdots \right]
\label{fourteen}
\een
Using (\ref{five}), the difference of areas in the large $\rho_0$ limit becomes
\ben
A_d(T) - A_d(0) = \frac{1}{2}\Omega_7 R^{7/2}r_H^7 \int_0^{\rho_0} d\rho\ \frac{\rho^{-7/2}}{(1+\frac{d^2}{\rho^2})^{25/4}} + \cdots 
\label{fifteen}
\een
The integral on the right hand side is finite in the limit of large $\rho_0$, so that we can replace the upper limit of integration by $\infty$. The leading result is then
\ben
A_d(T) - A_d(0) = C_0 \frac{\Omega_7 R^{7/2} r_H^7}{d^{5/2}} + \cdots \quad, C_0= \frac{2048}{69615}
\label{fifteen-1}
\een
Here the $\cdots$ represent subleading terms in the $r_H/d$ expansion.

As promised, the difference \eq{fifteen-1} is insensitive to the IR cutoff $\rho_0$.
The resulting difference of the entropies, to leading order,  using eq.(\ref{seere}), eq.(\ref{conj}), can be  expressed  as  
\ben
S(d,T)-S_{EE}(d,T=0) = C_0 \frac{\Omega_7 R^{7/2}r_H^7}{ 4 G_N d^{5/2}}
\label{sixteen}
\een

Before going on, let us mention one more way in which the dependence on $\rho_0$ in eq.(\ref{six}) can be made to cancel. 
Consider the supergravity background when the D0 branes are not at the origin of the Coulomb branch. In this case the Harmonic function in eq.(\ref{one})
is replaced by 
\ben
\label{harms}
H={R^7\over N} \sum_{i=1}^N {1\over | { \vec r} - {\vec r}_i |^7}
\een
where ${\vec r}_i$ is the location of the ith brane in the $9$ transverse directions. 
The area of the surface $x_1=d$ in this case is given by eq.(\ref{five}) with $H$ replaced by  eq.(\ref{harms}). 
Taking the difference of the Area in the geometry when the branes are at the origin of the coulomb branch and away from the origin then gives,
\ben
\label{diffe}
\Delta A= R^{7\over 2} \int dx_2 dx_3 \cdots dx_9 \big[{1\over r^{7/2}} - ({1\over N} \sum_i{1\over |\vec{r} - \vec{r_i} |^{7}})^{1/2} \big]
\een
Here $r$ is given in terms of $\rho$ by eq.(\ref{four}). At large $\rho$  the two terms in the brackets will  cancel to leading order. The second term  in the square brackets
due to the non-trivial Harmonic function can be expanded in a multipole expansion, the first correction to the leading term is due to the dipole and goes like 
${1 \over \rho^{9/2}}$, etc. The measure in the integral does like $\rho^7 d \rho$, so if the dipole term is present the integral will still blow up as 
$\rho\rightarrow \infty$. 
If fact one needs multipole contributions upto a fairly high order to vanish so that the leading contribution from the difference in the two terms in the bracket goes like 
${1\over \rho^{17/2 }}$.
While this is not elegant it can be arranged by choosing a suitable distribution of branes, and the resulting difference in area and hence entanglement entropies
 will  then be finite.

\subsection{Comparison with D0 brane Matrix theory}

In the proposal which follows we will identify each of the terms $S_{EE}(d,T=0)$ and 
$S(d,T)$ in (\ref{sixteen}) with quantities in the D0 brane matrix theory. However, as explained above, an unambigious comparison will be possible for the difference of these quantities with the difference of corresponding quantities in the D0 brane theory.
Using eq.(\ref{sixteen}), eq.(\ref{two}) and (\ref{eleven}) and the relation
\ben
\label{valGN}
G_N= 8\pi^6 g_s^2 l_s^8
\een
 this can be written as 
\ben
\label{finsee}
S(d,T)-S_{EE}(d,T=0)= B_0~ N^2 T_0^{14/5} d_0^{-5/2}
\een
where 
\begin{equation}
	B_0= \frac{480\ 2^{2/5} 15^{9/10} \pi^{13/2} \Gamma\left(\frac{5}{4}\right)}{49\ 7^{4/5} \Gamma\left(\frac{25}{4}\right)}
\end{equation}
We have defined
\ben
T = T_0 \Lambda
\label{eighteen}
\een
with 
\ben
\label{deflambda}
\Lambda={ (g_s N)^{1/3} \over l_s }
\een
and 
\ben
d = d_0 (g_s N)^{1/3} l_s
\label{nineteen}
\een

We aim to reproduce this behavior from the theory of D0 branes. The theory of D0 branes does not have any dimensionless parameter - there is only one scale which is the dimensional 't Hooft coupling $\lambda = g_{YM}^2N$. In terms of the bulk parameters this is given by
\ben
\label{defgym}g_{YM}^2 N = (g_s N)/l_s^3 = \Lambda^3. 
\een
This allows us to define a dimensionless temperature $T_0$ given in eq.(\ref{eighteen}). Also, 
the transverse radial coordinate $r$ is  proportional to the energy scale of the dual theory. This means that we should define a dimensionless distance $d_0$ 
as given by eq.(\ref{nineteen}). 
We note that the size of the ground state wave function in this system is also given by $(g_sN)^{1/3} l_s$, \cite{polchinskipeet,suss2}  and this is also the length scale $r_0$, eq.(\ref{condo}) beyond which the supergravity approximation breaks down; these observations agree with taking the dimensionless distance to be $d_0$ as above. 

From eq.(\ref{finsee}) we see that the  difference of the two bulk entanglement entropies when  expressed in terms of the appropriate dimensionless variables of the D0 brane matrix theory  scales like $N^2$. 
We also note that eq.(\ref{finsee}) is valid when eq.(\ref{condd}) holds, this condition can also be  expressed in terms of $d_0$ and $T_0$ and becomes,
\ben
\label{conde}
d_0\gg T_0^{2/5}
\een
Finally, eq.(\ref{finsee}) assumes that the supergravity approximation is valid, this requires, 
\ben
\label{condsugra}
T_0 \ll 1, N\gg 1
\een


It is also worth mentioning that from eq.(\ref{six})  and (\ref{seere})  it follows  that the  entanglement entropy itself  (obtained by ignoring  the $d$ dependent contributions) is given by 
\ben
\label{leads}
S_{EE} \sim N^2 ({\rho_0\over (g_s N)^{1/3} l_s})^{9/2}
\een
and also scales like $N^2$ when $\rho_0$ is expressed in terms of the appropriate dimensional length scale of the matrix theory. 

In the above discussion we have asserted that the entanglement entropy is proportional to area in Einstein frame. It is interesting to see what would happen if this was the area in string frame metric. In that case we get an answer
\ben
\Delta A_{\rm string-frame}= A_{\rm string-frame}(T)-A_{\rm string-frame}(T=0) \sim T_0^{14/5} d_0^{-13} l_s^8
\een
If we now use a UV cutoff which is the string length $l_s$ we see that $\Delta A_{\rm string-frame}/l_s^8$ can be again expressed in terms of quantities in the D0 brane quantum mechanics. Note however if this is taken to be a candidate for the entanglement entropy, the answer scales as $N^0$. This will not connect with the Bekenstein-Hawking formula and appears unnatural since the D0 brane theory has $N^2$ degrees of freedom. However we cannot rule out this possibility without a concrete calculation in the D0 brane quantum mechanics.

We now turn to a more detailed discussion of  D0 brane quantum mechanics. 


\section{Entanglement in D0 Brane Quantum Mechanics}
\label{d0qm}

Here we address the question: what is the bulk entanglement  in the dual description in terms of D0 brane quantum mechanics? 
Let us begin by reviewing some basics about the D0 brane matrix quantum mechanics. 
\subsection{Matrix Quantum Mechanics: Basic Facts}
The action for this $0+1$ dimensional supersymmetric Yang-Mills theory is given by
\ben
S = \frac{N}{2 (g_s N) l_s}{\rm Tr}\int dt  \left[ \sum_{I=1}^9 (D_tX^I)^2  - \frac{1}{ l_s^4} \sum_{I\neq J = 1}^9 [ X^I, X^J]^2 \right] + {\rm fermions}
\label{three-one}
\een
where $X^I(t)$ are $N \times N$ hermitian matrix functions of time and $D_t$ stands for the covariant derivative
\ben
D_t X^I = \partial_t X^I + i [A_t, X^I]
\label{three-two}
\een
This action has a $SU(N)$ gauge symmetry (actually the symmetry is $U(N)$, but the $U(1)$ decouples).
We can now fix a gauge $A_t = 0$. As usual, the resulting Gauss Law constraint imposes the condition that all physical states are invariant under a $SU(N)$ rotation
\footnote{More details of this model are discussed in Appendix B.}.
The hamiltonian in this gauge is 
\ben
H = \frac{1}{2}{\rm Tr}  \left[\frac {(g_s N) l_s}{N} \sum_{I=1}^9 (P^I)^2  + \frac{N}{(g_s N) l_s^5} \sum_{I\neq J = 1}^9 [ X^I, X^J]^2 \right] + {\rm fermions}
\label{three-three}
\een
where $P^I$ denote the conjugate momenta. 

This theory does not have any dimensionless parameter. This is seen clearly by rescaling 
\ben
X^I = (g_s N)^{1/3} l_s \tX^I~~~~~~~~~~~~~~P^I = \frac{1}{ (g_s N)^{1/3} l_s} \tP^I
\label{three-four}
\een
and the hamiltonian (\ref{three-three}) now becomes
\ben
H = \frac{(g_sN)^{1/3}}{2 l_s}{\rm Tr} \left[ \frac{1}{N} \sum_{I=1}^9 (\tP^I)^2  + N \sum_{I\neq J = 1}^9 [ \tX^I, \tX^J]^2 \right] + {\rm fermions}
\label{three-five}
\een
Thus the theory is characterized by a single energy scale
\ben
\Lambda = \frac{(g_s N)^{1/3}}{l_s}
\label{three-six}
\een
In this $A_t = 0$ gauge one is left with a time independent $SU(N)$ symmetry which also needs to be modded out. We will do this by diagonalizing one of the matrices, $X^1$. The remaining symmetry is now Weyl transformations which permute the eigenvalues of $X^1$ which we denote by $\lambda_i, i=1,\cdots N$, and mix up the matrix elements of the other eight matrices $X^I$ in a non-trivial fashion. In the following discussion we will ignore the fermions.

In the lowest energy state, all the nine matrices commute with each other. In this case all the matrices can be diagonalized simultaneously. If the eigenvalues are denoted by $x^I_i, i = 1 \cdots N$, these denote the locations of the $N$ D-branes. The origin of this Coulomb branch has $\lan X^I \ran \!=\! 0~~ -$ their dispersion provides the scale of the bound state, which is $(g_s N)^{1/3}l_s$. The supergravity description of this state is the $N$ coincident D0 brane solution discussed above. 

A generic state may be expressed in the form (the measure is derived in
\eq{measure} of Appendix B; in the following, we have omitted  the tilde sign
from $\tilde\Psi$ in Appendix B):
\ben
|\psi\ran =\int [d\mu] \Psi(\lambda_i; X^2_{ij},\cdots X^9_{ij})~ |\lambda_i; X^2_{ij},\cdots X^9_{ij} \ran +({\rm Weyl~ Transforms})
\label{three-seven}
\een
where we imposed the Weyl symmetry by summing over Weyl
transforms (according to \eq{permute-tilde-psi}). The measure is
\ben [d\mu] = \prod_{i=1}^N d\lambda_i
\prod_{I=2}^9[dX^I]
\een
Here $[dX^I]= \prod_i
dX_{ii}^I\ \prod_{i<j} dX^I_{ij}\ dX^I_{ji}$ is the standard Haar
measure. Here and in the following whenever we write $X^I$ the index $I$ runs
from 2 to 9.

Using the same basis, a generic operator may be expanded as 
\ben
\hO = \int [d\mu] \int [d\mu^\prime] \cO(\lambda_i, X^I_{ij} ; \lambda^\prime_i, X^{\prime~I})
 |\lambda_i; X^I\ran \lan\lambda^\prime_i, X^{\prime~I}| + {\rm Weyl~transforms}
\label{three-eight}
\een
In the low energy description, and at zero temperature, the space of eigenvalues $\lambda^i$ corresponds to one of the space directions, namely $x^1$, in 10 dimensional supergravity. 

\subsection{Target Space Entanglement Entropy\label{sec:target-EE}}
It is then clear that our calculation of the entanglement entropy across a $x^1=d$ surface  in a particular geometry in the bulk maps to 
a calculation of the {\it target space}  entanglement in the D0 brane quantum mechanics. That is we would like to restrict ourselves to the region $x_1>d$ and ask what are the operators we can have access to in this region; the von-Neumann entropy of the  density matrix  associated with this subalgebra of all observables 
is then the relevant  entanglement entropy. Compared to the single particle case briefly discussed in the introduction there are two extra features of this problem worth mentioning, both have to do with the fact that we are dealing with a system with many degrees of freedom.  

In general in a non-relativistic system with many particles, the analysis breaks up into different  sectors, each sector being specified by  which  of  the particles are present in the region of interest.  The corresponding set of operators in this sector correspond to all the measurements one can perform on these particles and the full sub algebra with which we associate the entropy is then a sum of the algebras of observables in each sector. 
In fact these sectors  are superselection sectors, since the observables in the algebra do not change the  particles in the region of interest. 

The second feature has to do with statistics. In our case the
different eigenvalues of the $X_1$ matrix correspond to fermion-like
degrees of freedom. More precisely the wave function $\Psi(\lambda_i,
X^I_{ij})$, eq.(\ref{three-seven}) has the property that it picks up a
minus sign under interchange of any given pair of indices
$i\leftrightarrow j$, i.e., under $\lambda_i
\leftrightarrow \lambda_j$, $X^{I}_{ii}\leftrightarrow
X^{I}_{jj}$ and $X^{I}_{ij}\leftrightarrow
X^{I}_{ji}$, the wave function, $\Psi \rightarrow -\Psi$.  This follows
from a special case of the general Weyl transformation
\eq{permute-tilde-psi} in Appendix B, where we choose to permute a given pair $(i,j)$. 
We are interested here in the target space region $x_1>d$.
The different super selection sectors are therefore specified only by the number of eigenvalues of $X_1$  meeting the condition $\lambda_i>d$, and not any particular choice of these eigenvalues. 

On general grounds, it then follows that the density matrix is block diagonal  in the different sectors and of the form
\ben
\label{deng}
\tilde\rho=\bigoplus_{m=0}^{m=N} \tilde\rho_m
\ee
where $\rho_{m}$ is the density matrix in the $m$-th sector in which
$m$ eigenvalues of $X_1$ meet the condition $\lambda>d$ (and the
remaining the $N-m$ eigenvalues are outside of this region). This is
similar to the equation \eq{rho-tilde-app} of Appendix A which
discusses the case of $N$ fermions; the notation $\tilde \rho_m$ here
is to be identified with $\tilde\rho_{m,N-m}$ of that equation.

Note that we can write \eq{deng} as
\ben
\label{bdrho}
\tilde\rho=\bigoplus_{m=0}^{m=N} p_m \ {\hat \rho}_m
\een
where $p_m$ is the probability to be in the $m$th 
\ben
\label{valpm}
p_m= Tr_m (\tilde \rho_m)
\een
 and ${\hat \rho}$ is the normalized density matrix in this sector, satisfying the relation
\ben
\label{broom}
Tr_m{\hat \rho}= 1
\een
The trace in eq.(\ref{valpm}), eq.(\ref{broom}) is restricted to the $m$th   sector.  Note in eq.(\ref{bdrho}) we have also allowed  for no eigenvalue  being  in the region of interest. 

It is worth emphasising that the probabilities $p_m$ satisfy the relation
\be
\label{relprob}
\sum_{m=0}^{m=N} p_m=1,
\ee
so that it  follows from eq.(\ref{valpm}) that  the full density matrix ${\tilde \rho}$ has the standard normalisation
\be
\label{stannorm}
Tr({\tilde \rho})=1.
\ee

Before proceeding let us also note that  the entanglement entropy, defined as
the von Neumann entropy of eq. \eq{bdrho} ({\it cf.} \eq{target-space-EE} of appendix A), is given by
\begin{eqnarray}
S_{EE}& = &  -\sum_m Tr_m \tilde\rho_m \ln (\tilde\rho_m)\label{entera}\\
 & = & - \sum_m p_m \ln p_m + \sum_{m=0}^N p_m Tr_m {\hat \rho} \ln ({\hat \rho})\label{enter}
\end{eqnarray}
 where the trace $Tr_m$ again denotes the trace within the sector with $m$ of the  eigenvalues lying in  the region of interest.
The  structure of the density matrix, eq.(\ref{bdrho} and entropy, eq.(\ref{enter}) are of the  general type which arises in  the presence of super selection sectors .   
And on  general grounds it follows that the distillable part of the entanglement is only the second term in eq.(\ref{enter}), while the first term $-\sum_mp_m \ln p_m$ is a classical piece which cannot be used as a quantum resource for teleportation, etc, \cite{ST}, \cite{V}.  




We also note that at non-zero temperature the relationship between the eigenvalues of the matrix $X^1$ and the coordinate $x_1$ in the background metric is not straightforward. However for regions far from the horizon these two quantities can be taken to be the same; thus since in this note we are dealing with the parametric region $d \gg r_H$ such an identification would be justified.


Let us now digress briefly to make one comment which is worth emphasising. 
For our proposal, that  the entanglement of some region in the  bulk  corresponds to target space entanglement in the boundary theory eq.(\ref{seere}),  to be sensible it is important that in a pure state the target space entanglement for a  region and its complement are equal. In the specific example we are considering this implies that  the target space entanglement corresponding to the region $x_1>d$ and $x_1<d$ are equal. 
It is easy to see that this is  the case and in fact  the reasoning we give below can be seen to apply immediately to a  general bulk region as well, so long  as this region can  be mapped suitably to a target space constraint in the boundary. 

Let us denote, for the discussion in this   paragraph only,  the density matrix for the region $x_1>d$ by ${\tilde \rho}^{(>d)}$. The corresponding entanglement entropy is given by 
\be
\label{eearg}
S_{EE}^{(>d)}= -Tr [{\tilde \rho}^{(>d)} \log {\tilde \rho}^{(>d)}].
\ee
Note that this trace has to be taken over  all the $N$ super selection sectors described above.  Now when  $m$ eigenvalues of $X^1$ take values in the range  $\lambda >d$,  $N-m$ eigenvalues lie in the complement, $\lambda <d$. Thus the $m^{\rm th}$  super selection for when we are considering the   $x_1>d$ region maps into the $(N-m)^{\rm th}$ sector  for the $x_1<d$ case.   A little more analysis also shows that the density matrices ${\tilde \rho}^{(>d)}_m$ for the $x_1>d$ 
region and correspondingly ${\tilde \rho}_{N-m}^{(<d)}$ for the $x_1<d$ region  make an equal contribution to their respective  entropies, $S_{EE}^{(>d)}, S_{EE}^{(<d)}$. 
This follows from the fact that in each sector the Hilbert space admits a tensor product decomposition, and the entanglement entropy for a pure state in a bipartite system is equal for both  of its constituent  Hilbert spaces \footnote{Actually to deal with the complication of fermion statistics  correctly we need to   embed the Hilbert space in each super selection sector in an extended Hilbert space where the required anti-symmetrisation constraint is not imposed. This extended Hilbert space admits a tensor product decomposition and that is enough to show the equality of the contributions ${\tilde \rho}^{>d}_m$ and ${\tilde \rho}_{N-m}^{<d}$ make to their respective entanglement entropies.}. The sectors where $m=0$ and $m=N$ have to be dealt with  as a special case. The equality  in these  sectors follows simply from the fact that the probability to not find any eigenvalue  taking values in the range $\lambda>d$ equals that to find all $N$ in $\lambda<d$ and vice-versa. 

 We now return to the main thread of our discussion. 
Consider  one of the terms in the state expressed in
(\ref{three-seven}), e.g. the first term. This has a given ordering of
the eigenvalue labels and the matrix elements of the remaining
$X^I$. The corresponding wavefunction is the probability amplitude
that the location of the $N$ D0 branes in the $x^1$ direction are
given by the $\lambda_i$. The diagonal matrix elements of $X^I$
represent open strings which begin and end on the same D0 brane, while
the off-diagonal matrix elements represent open strings which stretch
between different D0 branes. Now suppose in this particular term the
first $n$ eigenvalues have $\lambda_i > d$ while the rest have
$\lambda_i < d$. We will relabel the index $i$ for the latter set of
eigenvalues by the index $a$. We need to be able to perform
measurements which involve the matrix elements $X^I_{ij}, i,j =
1\cdots n, I = 2 \cdots 9$, while we do not wish to retain the
elements $X^I_{ab}, a,b = (n+1)\cdots N, I = 2 \cdots 9$. This leaves
us with the off-diagonal blocks $X^I_{ia}, X^I_{ai},\, i = 1\cdots n,
a = n+1 \cdots N, I = 2 \cdots 9$ and its transpose. Of course the
labelling of the matrix elements pertains to one specific term in the
sum (\ref{three-seven}). The question we are allowed to ask is {\em
  how many} eigenvalues are larger than $d$, not {\em which
  eigenvalues} are larger than $d$. The sum over Weyl transforms
precisely achieves this - ensuring that the D0 branes are identical
particles.

As explained below (see Section \ref{sec:two-proposals} as well as Appendix B), this leads us to two different proposals for the subalgebra of operators whose associated entanglement entropy corresponds to the quantity computed in the bulk. Let us first focus on the sector in which there are $n$ eigenvalues $\l_i$ in the region $x_1>d$. In our first proposal, the operator subalgebra relevant to this sector consists of operators in the Hilbert space of variables $\{\l_i, X^I_{ij}\}$  (see  Appendix A for a detailed discussion of operator subalgebras in the simple context of free fermions, e.g. \eq{o11}) which are of the form\\
\bea
\kern-60pt & \hO_n   = \tilde O_n \otimes \bar{\bf 1} + {\rm Weyl},\nonumber\\
& \kern-70pt  \tilde O_n = \int\  \prod_i \int_d^\infty d\l_i\,
\int_d^\infty d\l_i' \prod_{ij}^I dX^I_{ij} dX^{\prime I}_{ij} ~\tilde \cO_n (\{ \lambda_i ,\lambda^\prime_i\}; \{ X^I_{ij},  X^{\prime I}_{ij} \})
|\{\lambda_i, X^I_{ij} \}\ran
\lan \{\lambda^\prime_i, X^{\prime I}_{ij} \}|
+{\rm Weyl~Transf}
\nonumber\\
& \bar{\bf 1}= \int [d\mu_n] |\lambda_a, X^I_{ia} X^I_{ai} X^I_{ab} \}\ran
\lan \{\lambda_a, X^I_{ia} X^I_{ai} X^I_{ab} \}|
\nonumber\\
& \int [d\mu_n]  \equiv  \int_{-\infty}^d \prod_{a=n+1}^N\! d\lambda_a 
\int\! \prod_{a,b=n+1}^N \! [dX^I_{ab}]
\int \prod_{a=n+1}^N \prod_{i=1}^n [dX^I_{ia} dX^I_{ai}]
\label{three-nine}
\eea
The full operator algebra consists of contribution of operators from the various $n$-sectors.

From the above definition, it is clear that in this proposal we are
tracing over not only the $(N-n) \times (N-n)$ block of the matrices
$X^I$, but also the off-diagonal blocks. This means that we are not
performing measurements on the open strings which join the D0 branes
in the $x_1 > d$ region with those in the $x_1 < d$ region.

In our second proposal the subalgebra of operators in the $n$-sector
consists of operators in a Hilbert space of coordinates $\{\l_i, X^I_{ij},
X^I_{ia}\}$, and are of the form
\bea
\kern-60pt & \hO_n   = \tilde O_n \otimes \bar{\bf 1} + {\rm Weyl},\nonumber\\
& \kern-60pt  \tilde O_n =\prod_i\  \int_d^\infty  d\l_i\,
\int_d^\infty d\l_i' \prod_{ij}^I \int\ dX^I_{ij} dX^{\prime I}_{ij} 
\prod_{ia,ai}^I \int\ dX^I_{ia} dX^{\prime I}_{ia} dX^I_{ai} dX^{\prime I}_{ai}\times
\nonumber\\
& \kern-30pt \tilde \cO_n (\{ \lambda_i ,\lambda^\prime_i\}; \{ X^I_{ij},  X^{\prime I}_{ij}; X^I_{ia},  X^{\prime I}_{ia};
X^I_{ia},  X^{\prime I}_{ia} \})
|\{\lambda_i, X^I_{ij}, X^I_{ia} \}\ran
\lan \{\lambda^\prime_i, X^{\prime I}_{ij}, X^{\prime I}_{ia} \}|
+{\rm Weyl~Trans}
\nonumber\\
& \bar{\bf 1}= \int [d\nu_n] |\lambda_a, X^I_{ab} \}\ran
\lan \{\lambda_a, X^I_{ab} \}|
\nonumber\\
& \int [d\nu_n]  \equiv \int_{-\infty}^d\! \prod_{a=n+1}^N\! d\lambda_a 
\int \prod_{a,b=n+1}^N\! [dX^I_{ab}]
  \label{three-12}
\eea
In this proposal we are tracing over only the $(N-n) \times (N-n)$ block of the matrices $X^I$. This means that our measurements include those made on open strings which join the D0 branes in the $x_1 > d$ region with those in the $x_1 < d$ region.

In this paper we will largely focus on the bosonic degrees of freedom in the quantum mechanics and not discuss the fermionic ones. However let us at least mention that 
the fermionic degrees of freedom $\theta_A$ which are also $N\times N$ matrices must  be dealt  with in the same way as the $X^I, I=2, \cdots 9$ matrices. 
This means in the first proposal we only retain the $(\theta_A)_{ij}$ blocks and trace over the $(N-n) \times (N-n)$ blocks $(\theta_A)_{ab}$, as well as the off-diagonal blocks, 
$(\theta_A)_{ai}, (\theta_A)_{ia}$. 
In the second proposal we retain the $(\theta_A)_{ij}$ and the  $(\theta_A)_{ai}, (\theta_A)_{ia}$ blocks and only trace over the $(\theta_A)_{ab}$ block. 

For a given state one can now compute the reduced density matrix which correctly reproduces expectation values of either of the set of operators, and from this the von Neumann entropy. The formalism to write this down is explained in the Appendix and will also be elaborated in the next subsection.  Our conjecture is that one of these will correspond to the bulk entanglement entropy computed in section {\ref{d0sugra}). 

\subsection{The two proposals for target space EE\label{sec:two-proposals}}
  
Before proceeding,  some more comments are worth making at this stage.  We note  that  some  motivation for the two proposals  above come from the Coulomb branch solutions. In supergravity it is known that there are solutions in which the D0 branes are displaced from the origin and the harmonic function takes the more general form eq.(\ref{harms}), with  $\vec{r}_i$ specifying the location of the ith  brane. These solutions also correspond to bound states at zero energy in the matrix  theory.  Consider such a solution in which the $x_1$  coordinate of some of  the  D0 branes lies in the region $x_1<d$, i.e. outside the region of interest.  In defining the entanglement 
  if these branes are to be excluded, then   the open strings 
stretching between these excluded branes should also be dropped. This still leaves the choice of whether the degrees of freedom corresponding to the open strings stretching between the branes inside the region,  with  $x_1>d$, and those outside,   with $x_1<d$, should be  retained or dropped.  Correspondingly, in the density matrix of  the matrix theory we have two choices of retaining the off-diagonal degrees of freedom stretching between the eigenvalues with $\lambda_i<d$ and $\lambda_i>d$, as discussed above. See figure \ref{fig-coulomb-branch}. 
For some more details, see Appendix B (section \ref{target-EE-matrices}).

\begin{figure}
\begin{center}
\hspace{5ex}\includegraphics[scale=.3]{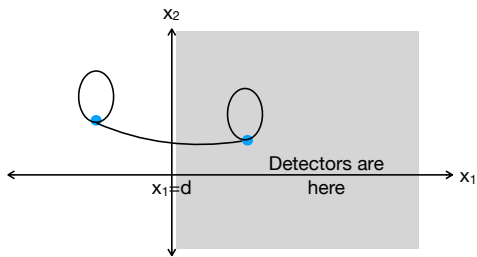}\\
\caption{A typical configuration for the case $N=D=2$, where we have
  two $2\times 2$ matrices $X= {\rm diag}[\l_1, \l_2]$, $Y= {\rm
    diag}[y_{11}, y_{22}]$; $(\l_i, y_{ii})$ represent the coordinates
  of the two D0 branes, $i=1,2$. The figure depicts the situation in
  which one of the D0 branes, say with coordinates ${\bf
    x_1}=(\vev{\lambda_1}, \vev{y_{11}})$, is in region $A:x^1>d$,
  i,e. $\lambda_1 >d$, while the other D0 brane with coordinates ${\bf
    x}_2=(\vev{\lambda_2}, \vev{y_{22}})$ is in $\bA: x^1<d$,
  i.e. $\lambda_2 <d$. The {\it variables} $\l_1, y_{11}$ represent
  an open  string beginning and ending on 
  the first D0-brane; they are in the region of interest $A$ and hence
  must be included in the operator algebra. Similarly the 
  {\it variables} $\l_2, y_{22}$ represent an open string beginning and
  ending on the second D0-brane; they are in the region of interest
  $\bA$ and hence should be excluded from the operator algebra.
  $y_{12}, y_{21}$ represent open strings straddling between regions
  $A$ and $\bA$. One might wish to exclude the $y_{12}, y_{21}$ from
  the operator algebra (first proposal), or include them (second
  proposal). Note that, the two D0 branes are actually
  indistinguishable; as \eq{weyl},\eq{permute-tilde-psi} indicate, the
  situation described above is indistinguishable from the one in which
  the D0 branes are interchanged; hence the above definitions have to
  take that into account, as was done for case of $N$ fermions in
  Appendix A (section \ref{appendix}).}
\label{fig-coulomb-branch}
\end{center}
\end{figure}

It is worthwhile to emphasize that even though we draw motivation from a generic point on the Coulomb branch, the state we are discussing is at the origin of the Coulomb branch. The supergravity solution for this is a set of coincident D0 branes. However, in the matrix quantum mechanics this state has a non-trivial wavefunction which has a spread of $\sim N^{1/3}l_s$. This means that while the {\em expectation value} of the matrices vanish in this state, there is a non-zero probability amplitude (described by the wavefunction)  for having a configuration described by values of $\lambda_i, \lambda_a, X^I_{ij}, X^I_{aj}, X^I_{ab}$, using the notation described above. The open strings we refer to above are simply a description of this kind of configuration. These comments are also true for a state of the kind in eq.(\ref{harms}) where the branes are displaced from the origin but continue to lie in the region of validity of the supergravity approximation; while the expectation value of the matrices do not vanish now, there is a non-zero probability amplitude for various values of $\lambda_i, X^{I}_{ij},$ etc. 

In either of the above proposals, there are $O(N^2)$ degrees of freedom which are traced out. It is therefore natural to expect that the entanglement entropy will be proportional to $N^2$. We note that the fact there are $O(N)$ sectors in the sum, eq.(\ref{enter}) does not alter this estimate. 
If $O(N)$ sectors contribute and generically each sector gives a contribution of $O(N^2)$ by which we mean that the normalized density matrix ${\hat \rho}_m$ has 
$Tr_m {\hat \rho}_m \ln ({\hat \rho}_m)\sim O(N^2)$, then the final result for the second term in eq.(\ref{enter}) would still be $O(N^2)$. 
The first term in eq.(\ref{enter}) which is the classical piece is much smaller and can at most be $O(\ln N) $.

When the bulk is a black hole we should consider the D0 brane quantum mechanics in a thermal state with the same temperature $T$. There are now two dimensionful quantities in the calculation. The first is the temperature $T$ and the second is the value of $d$ which has been used to define the subalgebra.  It is clear from the discussion from equation (\ref{three-one}) to (\ref{three-six}) that the energy scale in the 't Hooft limit is given by $\Lambda$ defined in (\ref{three-six}), while the scale which relates supergravity distances with the eigenvalues is $(g_s N)^{1/3} l_s$. Therefore the D0 brane quantum mechanics answer for the entanglement entropies will involve the dimensionless temperature $T_0 = T/\Lambda$ and the dimensionless $d_0$ introduced above. Once this is done, the answer should be simply proportional to $N^2$, exactly as in the supergravity calculation. As explained in the previous section,  to keep the bulk calculation within the realm of the supergravity approximation one could compare  the difference of the entropies at finite and zero temperature, this would allow for a  precise test of the coefficient in the area term in eq.(\ref{seere}).

While a bound state which corresponds to $N$ D0 branes has been shown to exist \cite{bound}, an explicit analytic form is not known. This makes an analytic check of our proposal difficult. It should be, however, possible to express the target space entanglement entropy discussed above in a path integral formulation : then numerical calculations along the lines of \cite{Hanada:2016zxj} can be used to provide a concrete check of our proposal.

Before ending this subsection let us also mention that a useful toy model to understand target space entanglement is to consider the case of a single bosonic matrix quantum mechanics with no external potential. In this case the additional $X^I$ are not present and we only have the eigenvalues $\lambda_i$. As is well known the $\lambda_i$ can be considered as the coordinates of $N$ free fermions moving on a line. The above description of the relevant subalgebra of operators is in a first quantized description. In a second quantized description, the Hilbert space becomes a product. The subalgebra of operators pertaining to the subregion $\lambda > 0$ are given by $M$ body operators of the form
\bea
{\cal{F}} = 
& \int_0^\infty \prod_{i=1}^M [d\lambda_i d\lambda_i^\prime]& ~\psi^\dagger (\lambda_1)
\psi^\dagger (\lambda_2) \cdots \psi^\dagger (\lambda_M) \nonumber \\
& & F_M(\lambda_1 \cdots \lambda_M; \lambda_1^\prime \cdots \lambda_M^\prime)~\psi (\lambda_1^\prime)\psi(\lambda_2^\prime)\cdots \psi(\lambda_M^\prime)
\label{three-13}
\eea
where $\psi(x),\psi^\dagger(x)$ are the second quantized fermion fields.
In the sector where there are $n$ particles in this region the operators which have nonzero expectation values must have $M \leq n$. It can be then shown easily that the functions 
$F_M$ are in one-to-one correspondence with the matrix elements of operators in the first quantized description in the sector where there are $M$ particles in the sub-region.
In fact, for free fermions one may use well known methods to compute the reduced density matrix \cite{chuerta} to show that the density matrix obtained in the second quantized description is exactly the same as the first quantized description discussed above. Details of this are provided in appendix A.

\subsection{The Sector-wise entanglement}
Consider the sector where $n$ eigenvalues  satisfy the condition $\lambda_i>d$.  Let us  use the Weyl symmetry and arrange for   these to be the first $n$ eigenvalues of $X_1$. Then in the first proposal we would also retain in the density matrix the $X_{ij}, (i,j\le n)$ degrees of freedom and 
``integrate out" everything else. Starting with a wave function $\Psi(\lambda_i, X^I_{ij})$ with unit norm,
\ben
\label{norms}
\int d\lambda_i DX_{ij}^I |\Psi( \lambda_i, X^I_{ij}) |^2=1,
\een
where in the integral $\lambda_i\in [-\infty,\infty]$ and the measure for $X_{ij}^I$ is the standard flat measure for Hermitian matrices, with a range as explained in the appendices A,B.
We then get that the density matrix in this sector is given by ({cf.} \eq{deng})
\ben
\label{denmatra}
\tilde\rho_n(\lambda_i, X^I_{ij}; \lambda'_{i'}, X^{'I}_{i'j'})={ N \choose n}  \int D\chi_A   \Psi^*(\lambda_i, X^I_{ij}, \chi_A) \Psi(\lambda'_{i'}, X^{'I}_{i'j'}, \chi_A)
\een
To save clutter we have denoted all variables to be integrated over generically as $\chi_A$. These include $\lambda_i, i>n$ and $ X^{I}_{ij}$, where one or both indices $i,j$ are greater than $n$. Note that the range of integration over these variables is as follows $\lambda_i, i>n$ take values $ \in [-\infty, d]$, while $X^I_{ij}$, with $i$ or $j$ $>n$,  are  to be integrated over their  full range  (real line for $i=j$ and complex plane for $i\ne j$). It is important to note that  the variables being integrated out, $\chi_a$, appear in both $\Psi^*$ and $\Psi$. The combinatorial factor ${N \choose n}$ arises as follows. 
The case with $n$ eigenvalues of $X_1$ being greater than $d$ can  arise in ${N \choose n}$ different ways, by the fermionic symmetry these all give the same contribution to the density matrix resulting in this combinatorial factor. 

Note that the density matrix $\rho_n$ is  an operator in the space of the  degrees of freedom that remain after imposing the target space constraint and once it is known we can in principle calculate its contribution to the entropy,  $Tr_n \rho_n \ln (\rho_n)$. Summing the contributions from the different sectors then gives the full entanglement entropy, eq.(\ref{entera}). 

 In the second proposal after arranging for the first $n$ eigenvalues to be greater than $d$ we retain :  $\lambda_i, i\le n$, $X_{ij}, i,j<n$. In addition we retain  the  degrees of freedom, $X^I_{a,i}, X^I_{ia}, $ with, $i<n, a>n$;  these satisfy the relation  $X^I_{ai}=(X^I_{ia})^*$. The density matrix now depends on these degrees of freedom as well, and eq.(\ref{denmatra})  is replaced by 
 \ben
 \label{dent}
\kern-60pt \tilde\rho_n(\lambda_i, X^I_{ij}, X^{I}_{ai}; \lambda'_{i'}, X^{'I}_{i'j'}, X^{I}_{a',i'})= {N \choose n} \int D\chi_A \Psi^*(\lambda_i,  X^I_{ij}, X^I_{ai}, \chi_A ) \Psi(\lambda_{i'}, X^I_{i'j'}, X^I_{a'i'}, \chi_A)
 \een
 where now the $\chi_a$ variables include :  $\lambda_i, i>n$, $X^{I}_{ij}, i,j>n$. 
 The range of integration for these variables are as above and the combinatorial factor has the same origin as in the previous case. 
 More details can be found in Appendix A and B.

 While  we have not been explicit about fermionic degrees of freedom here  they are to be included in a manner analogous to the $X^I$ degrees of freedom, as was discussed after eq.(\ref{three-12}) above. 
 
 Finally let us note the form  for $\rho_n$  if we start not with a wave function $\Psi$ but with a density matrix for the full system, as would be the case when we consider the finite temperature case where 
 \ben
 \label{denim}
 \rho= {e^{-H/T} \over \sum_i e^{-H/T}},
 \een
 where $H$ is the Hamiltonian and the index  $i$ denotes sum over all states. 
 The density matrix can now be regarded as a  general  function $\rho(\lambda_i, X^I_{ij}; \lambda_{i'}, X^I_{i'j'})$, with $i,j, i',j',$ taking values $1, 2, \cdots N$. 
 
 In this case similar reasoning as above shows that  for the first proposal  eq.(\ref{denmatra}) is  replaced by 
 \ben
 \label{fTdenm} 
 \tilde\rho_n(\lambda_i, X^I_{ij}; \lambda'_j, X^{'I}_{i'j'})={ N \choose n}  \int D\chi_A   \rho(\lambda_i, X^I_{ij}, \chi_A; \lambda'_{i'}, X^{'I}_{i'j'}, \chi_A)
 \een
 where $\chi_a$ as above denotes the variables, $\lambda_i, i>n$ and  $X^I_{ij}$ ,  where one or both labels, $i,j>n$.
 Whereas in the second proposal eq.(\ref{dent}) is replaced by 
 \ben
 \label{dent2}
\kern-40pt \tilde\rho_n(\lambda_i, X^I_{ij}, X^{I}_{ai}; \lambda'_j, X^{'I}_{i'j'}, X^{'I}_{a',i'})= {N \choose n} \int D\chi_A \rho(\lambda_i,  X^I_{ij}, X^I_{ai}, \chi_A ;\lambda_{i'}, X^{'I}_{i'j'}, 
 X^{'I}_{a'i'}, \chi_A)
 \een
 where $\chi_A$ now includes, $\lambda_i, i>n$, and $X^I_{ij}$, with both $ i,j,>n$.

\section{Dp Branes ($p < 3$)}
\label{dpsugra}

The results of section (\ref{d0sugra} generalize to Dp branes with $p < 3$. The string frame metric and the dilaton for the near horizon geometry of $N$ coincident near-extremal black Dp branes are
\bea
ds^2 & = & \left(\frac{R}{r} \right)^{-n/2} \left[- f(r) dt^2 + dy_1^2 + \cdots dy_p^2 \right] +  \left(\frac{R}{r} \right)^{n/2} \left[ \frac{dr^2}{f(r)} + r^2 d\Omega_{n+1}^2 \right] \nonumber \\
e^{-\phi/2} & = & \left(\frac{R}{r} \right)^{\frac{n(p-3)}{8}} 
\label{two-one}
\eea
where
\bea
n & = & 7-p~~~~~~~~R^n = (4\pi)^{(n-2)/2} \Gamma(n/2) l_s^n (g_s N) \nonumber \\
r^2 & = & x_1^2 + \cdots x_{9-p}^2 = x_1^2 + \rho^2 \nonumber \\
f(r) & = & 1-\left(\frac{r_H}{r} \right)^{n}
\label{two-two}
\eea
and the temperature is given by
\ben
T = \frac{n}{4\pi R} \left(\frac{r_H}{R}\right)^{\frac{n-2}{2}}
\label{two-three}
\een
The brane directions $y_i$ each have an extent $L$.
Consider once again a $x_1=d$ surface where $d > r_H$. The Einstein frame area of this surface is given by
\ben
A_d(T) = \Omega_{n}R^{n/2}L^p \int_0^\infty d\rho \frac{\rho^{n/2}}{(1+\frac{d^2}{\rho^2})^{n/4}}\left[ 1 + \frac{r_H^n}{\rho^n} \frac{1}{(1+\frac{d^2}{\rho^2})^{\frac{n}{2}+1}}\right]^{1/2}
\label{two-four}
\een
This integral is divergent at the upper limit. However, as in the case of zero branes, the difference $A_d(T)-A_d(0)$ is finite.
Performing a low temperature expansion as in the previous section we obtain the difference of the areas which is once again insensitive to the IR cutoff on $\rho$ and the entropy difference is then given by
\ben
\Delta S_{EE}= \frac{\Omega_n \Gamma \left(\frac{n-2}{4}\right) \Gamma \left(\frac{n+3}{2}\right)}{4\Gamma \left(\frac{3 n}{4}+1\right)} \frac{L^p R^{n/2}r_H^n}{\epsilon^8 d^{\frac{n}{2}-1}}
\label{two-five}
\een
Using the expression for $R$ in (\ref{two-two}) and $r_H$ in terms of the temperature in (\ref{two-three}) we get
\ben
\Delta S_{EE} = C_p \frac{(g_s N)^2 l_s^8}{\epsilon^8}~(g_s N)^{\frac{6-n}{2(n-2)}}~l_s^{\frac{3n^2-18n+32}{2(n-2)}}~T^{\frac{2n}{n-2}}~L^{7-n}~d^{1-\frac{n}{2}}
\label{two-six}
\een
where
\begin{equation}
	C_p = (n+1) ~2^{\frac{3 n^2- 4n+ 12}{2(n-2)}} ~\pi^{\frac{n(5n-2)}{4(n-2)}} ~n^{-\frac{2n}{n-2}}~ \Gamma\left(\frac{n}{2}\right)^{\frac{3n-2}{2(n-2)}} ~\frac{\Gamma\left(\frac{n-2}{4}\right) }{\Gamma\left(\frac{3 n}{4}+1\right)}
\end{equation}
We now need to express the temperature, the p-brane extent and the quantity $d$ in terms of their appropriate scales. The energy scale $\Lambda$ of the Dp brane theory is provided by the 't Hooft coupling
\ben
g_{YM}^2 N = \frac{(g_s N)}{l_s^{n-4}} \Rightarrow \Lambda = (g_s N)^{\frac{1}{n-4}}l_s^{-1}
\label{two-seven}
\een
This means that we need to express $T$ and the extent $L$ in these units,
\ben
T = T_0 \Lambda~~~~~~~~~L = L_0 \Lambda^{-1}
\label{two-eight}
\een
The transverse distance in the geometry is, however proportional to this energy scale multiplied by $l_s^2$. This means that we need to express
\ben
d = d_0 \Lambda l_s^2
\label{two-nine}
\een
Once again, when expressed in terms of these dimensionless quantities, the result should not involve $g_s$. This can happen only if the UV cutoff $\epsilon$ is proportional to the 10 dimensional Planck scale. Using this cutoff, we are left with a final answer proportional to $N^2$, 
\ben
\Delta S_{EE} = B_p~ N^2~T_0^{\frac{2n}{n-2}}~L_0^{7-n}~d_0^{1-\frac{n}{2}}
\label{two-ten}
\een
where
\begin{equation}
	B_p = (n+1) ~2^{\frac{3 n^2- 14n+ 32}{2(n-2)}} ~\pi^{\frac{5 n^2 -26n +48}{4(n-2)}} ~n^{-\frac{2n}{n-2}}~ \Gamma\left(\frac{n}{2}\right)^{\frac{3n-2}{2(n-2)}} ~\frac{\Gamma\left(\frac{n-2}{4}\right) }{\Gamma\left(\frac{3 n}{4}+1\right)}
\end{equation}

\section{Entanglement in $Dp$ brane field theory}

The discussion of a candidate subalgebra of operators in the $SU(N)$ Yang-Mills theory living on the Dp brane (for $p < 3$) worldvolume is completely analogous to that for D0 brane quantum mechanics. The matrices are now functions of the spatial coordinates on the Dp brane worldvolume $\xi$. The bosonic fields are now worldvolume gauge fields $A_\mu (\xi), \mu = 1 \cdots (p+1)$ and the transverse Higgs fields $X^I (\xi)$ with $I=1,\cdots 9-p$. We then work in a gauge where one of these Higgs fields, $X^1$ is chosen to be diagonal with elements $\lambda_i (\xi)$ and consider a division of the space of $\lambda (\xi)$ into two parts, corresponding to $\lambda_i (\xi) > d$ and $\lambda_i (\xi) < d$ . As in section (\ref{d0qm}), there are two choices for the corresponding operator sub-algebra. The generalization for the choice (\ref{three-nine}) involves an expression 
\bea
\kern-60pt & \hO_n   = \tilde O_n \otimes \bar{\bf 1} + {\rm Weyl},\nonumber\\
& \kern-50pt  \tilde O_n = \prod_\xi  \prod_i \int_d^\infty d\l_i(\xi)\,
\int_d^\infty d\l_i'(\xi) \prod_{ij}^I dX^I_{ij}(\xi)
dX^{\prime I}_{ij}(\xi)\times\nonumber\\
& \tilde \cO_n (\{ \lambda_i(\xi) ,
\lambda^\prime_i(\xi)\}; \{ X^I_{ij}(\xi),  X^{\prime I}_{ij}(\xi) \})
|\{\lambda_i(\xi), X^I_{ij}(\xi) \}\ran
\lan \{\lambda^\prime_i(\xi), X^{\prime I}_{ij}(\xi) \}|
+{\rm Weyl~Trans}
\nonumber\\
& \bar{\bf 1}= \int [d\mu_n] |\lambda_a, X^I_{ia} X^I_{ai} X^I_{ab} \}\ran
\lan \{\lambda_a, X^I_{ia} X^I_{ai} X^I_{ab} \}|
\nonumber\\
&\int [d\mu_n]  \equiv \prod_\xi \int_{-\infty}^d \prod_{a=n+1}^N\!
d\lambda_a(\xi) 
\int\! \prod_{a,b=n+1}^N \! [dX^I_{ab}(\xi)]
\int \prod_{a=n+1}^N \prod_{i=1}^n [dX^I_{ia}(\xi) dX^I_{ai}(\xi)]
\label{three-nine-a}
\eea
This equation should be regarded in the same spirit as 
\eq{three-nine}; the operator $\cO (\{ \lambda_i (\xi) ,\lambda^\prime_i (\xi) \}; \{ A^\mu_{ij} (\xi) X^I_{ij} (\xi);  A^{\prime \mu}_{ij} (\xi) X^{\prime I}_{ij} (\xi) \})$ belongs to the Hilbert space of the
variables $\{ \lambda_i (\xi) , A^\mu_{ij} (\xi) X^I_{ij} (\xi) \}$. 
The measure here is again a generalization of (\ref{three-nine}) with the additional terms involving the gauge fields and the integrals replaced by functional integrals. 
Note that  (\ref{three-nine-a}) involves integration over functions, and the restrictions on the ranges of integration are over the values of the function {\it at each point on the base space $\xi$}.
The subalgebra of operators for our second proposal also follows in a similar fashion.

\section{Discussion}
In this paper we explored the idea that  in any smooth spacetime, to leading order, the Bekenstein bound is saturated, eq.(\ref{conj}), leading to the proposal that for a pure state the entanglement of any co-dimension one region is given by the area of its boundary in units of $G_N$, eq.(\ref{seere}). We have shown that for a special choice of bulk regions the bulk entanglement can be mapped, upto one ambiguity, to the target space entanglement in the boundary theory. Our proposal  can therefore be tested precisely using numerical calculations along the lines of \cite{Hanada:2016zxj}. If our proposal lives up to precise tests, this would mean that the UV cutoff which makes the entanglement entropy finite  in string theory is the Newton constant, and not the string length. In fact, this is the lesson from the $c=1$ example in \cite{Das:1995vj, Hartnoll:2015fca}.

We have described the bulk region of interest in a coordinate system and used the relationship between bulk coordinates and target space of the matrix theory in this coordinate system. The notion of the region itself and its bounding surface is of course coordinate invariant. In a different coordinate system the map to matrix theory target space will be different, and therefore the target space restriction will be different. The result, however, will remain the same.

One would like to extend our considerations to more general regions in the bulk. A  preliminary study suggests that  this might be possible. For example in the D0 brane case consider a spherical bulk region given by,
\ben
\label{mcondo}
\sum_{i=1}^9 (x^i)^2 \le  R^2
\een
In the matrix  theory  the corresponding operator $\sum Tr ({\hat X}^i)^2 $ is Hermitian and one can choose a gauge where it is diagonalized\footnote{We are grateful to Shiraz Minwalla for a discussion on this point.}. This suggests that our considerations might be  extendable to more general bulk regions as well. 
Such an extension would be particularly interesting for a region of the type eq.(\ref{mcondo}), since by changing the radius one could then deform the bulk region smoothly from being away from the black hole horizon to lying on it. It is also worth mentioning that if our proposal is correct the entropy contained in the region
eq.(\ref{mcondo}) is temperature independent, since its area is independent of $T$, as can be easily seem from eq.(\ref{ten}). The thermal and entanglement contributions to the entropy presumably  trade-off against each other keeping the total unchanged. 

It is also worth commenting that various positivity properties, e.g. positivity of relative entropy and mutual information,  \cite{OP}, \cite{Wittenxx}  should  hold for target space entanglement. For example, positivity of relative entropy and its monotonicity under inclusion of algebras are general properties which should also apply to target space entanglement;  from these follow positivity of mutual information and strong subadditivity, etc. Using eq.(\ref{seere}) these properties can be mapped to properties of areas bounding regions in the bulk. 
A preliminary analysis suggests that they are  true and in some cases the inequalities are in fact  saturated. For example, consider two target space regions, $A: d<x_1$ and $B: d_2<x_1<d$, with $A\cup B: d_2<x_1$. Then it is manifestly true that their mutual information $I(A,B)$ is positive, since, 
\ben
\label{mine}
I(A,B)=S(A)+S(B)-S(A\cup B) =2 {A(x_1=d)\over 4 G_N}>0
\een
Similarly considering two overlapping regions with $A \cap B \ne 0$, it is easy to  see that the strong subadditivity condition would be saturated
\ben
\label{ssa}
S(A)+S(B)=S(A\cup B) + S(A \cap B).
\een
See Figure \ref{fig-venn}.

\begin{figure}
  \begin{center}
  \includegraphics[scale=0.3]{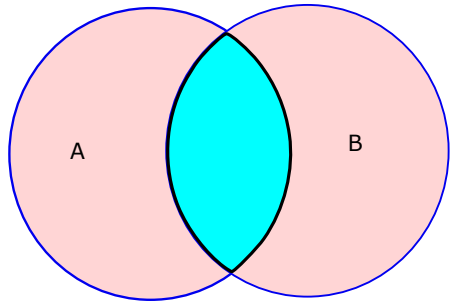}
    \caption{Strong subadditivity: consider regions $A$ and $B$. The
      set $A\cup B$ is depicted in pink, whereas $A\cap B$ is depicted
      in blue. Denoting the area of the boundary of regions $A$, $B$
      etc. as $a(A), a(B)$ etc, it is straightforward to see that
      these areas satisfy the equality $a(A)+ a(B) = a(A\cup B) +
      a(A\cap B)$.}
  \label{fig-venn}
  \end{center}
  \end{figure}

As discussed in the introduction, for field theories, where the
degrees of freedom live in both spatially extended regions and time,
one can consider a more general notion of entanglement which arises
when we consider observables which only access both a spatially
localized region and a restricted region in target space.  It would be
worth exploring this more general notion in the context of AdS/CFT
further. Without target space restrictions the bulk dual of the
boundary entanglement entropy is the Ryu-Takayanagi surface. With only
target space restrictions and no any restrictions along the spatial
directions, we have proposed here, for some cases, that the target
space entanglement maps to bulk entanglement of an appropriate bulk
region. The more general notion combining both spatial and target
space restrictions would then interpolate between these two and it
will be interesting to understand its bulk dual in more detail. While
the Ryu-Takayanagi surface is extremal, our preliminary
considerations here suggest that more generally when target space
constraints are also included, the bulk surface is not extremal and in
fact could be of a quite general type.

We find it very  interesting that even for the  restricted kind of spatial regions considered here, a precise map of bulk entanglement exists in the boundary theory.
Since the notion of bulk locality is not precise in a theory of gravity, this was not a priori clear.  The boundary theory of course exists for all value of the coupling and all values of $N$ (in terms of the quantities appearing in eq.(\ref{finsee}) all values of $T_0,d_0,N$. Thus one could consider how the target space entanglement changes as one goes to weak coupling, and smaller values of $N$. The $\alpha'$ and string loop corrections correspondingly become important in the bulk, and bulk locality would become  a more imprecise notion, but the target space entanglement  would continue to be well defined. One could also  try to check this by computing $\alpha'$ corrections in the bulk.

It is clearly important to find additional, and more doable, tests for our conjecture, eq.(\ref{conj}), eq.(\ref{seere}). One possibility might be to try and investigate this in a semi-classical path integral which attempts to implement the replica trick in the bulk, about a smooth spacetime background \footnote{We are grateful to Shiraz Minwalla for this suggestion.}. 

We end by noting that if, as our preliminary investigation here suggests, the notions of target space entanglement along with  its generalization mentioned above  which combines spatial and target space constraints,
can provide a precise notion of bulk entanglement,  they  would clearly be important for studies related to  information loss and more generally black hole physics.

\section{A personal note from S.R.D.} I came to know Peter Freund closely during my years as a graduate student at University of Chicago, and we remained in touch ever since. His original style of doing physics has been a major influence in my life, and his enthusiasm has been contagious. I am honored to contribute this work to his memorial volume.

\section{Acknowledgements}

We thank Shaun Hampton, Antal Jevicki, Sinong Liu, Shiraz Minwalla, Suvrat Raju and Ashoke Sen for discussions. S.R.D would like to thank Tata Institute of Fundamental Research for hospitality during numerous extended visits over the years which led to this work. The work of S.R.D is partially supported by National Science Foundation grant NSF/PHY-1818878.
A.K., G. M. and S. P. T. acknowledge   the support of the Govt. Of India, Department of Atomic Energy, under Project No. 12-R\&D-TFR-5.02-0200 and support from the Quantum Space-Time Endowment of the Infosys Science Foundation. S. P. T. acknowledges support from a J. C. Bose Fellowship, Department of Science and Technology, Govt. of India.

\section{Appendix A: Target Space Entanglement Entropy}
\label{appendix}
In this appendix we first present the formalism of target space entanglement entropy in the context of non-relativistic quantum mechanics of N fermions. Then we go on to prove the equivalence of reduced density matrix constructed by particle number sector (and the consequent EE) in the first quantized formulation with the standard second quantized theory.

We follow \cite{DMT:2018} \cite{Mazenc:2019ety} to define the EE in the target space. The key idea here is the algebraic definition of EE which relies on the usage of a theorem (Artin-Wedderburn) which says that given any algebra, there always exists a decomposition of the Hilbert space with the structure of direct sum over tensor products. Once restricted to a particular sector, one can use the usual notions of reduced density matrix due to the tensor product structure. However here we do not distinguish between "classical" and "quantum" contribution to the EE. 

We introduce the notion of target space EE for a system of $N$
fermions moving on the real line $R$; from the first quantized
viewpoint, $R$ is the ``target space''.  We would like to define the
EE of a target space subregion region $A\subset R$, e.g. $A$ could be
the region $x>d$ for some real number $d$. Given such a region and its
complement $\bA$, the one-particle Hilbert space, $\mH_1$ has the
structure of a direct {\it sum}, rather than product, of the form
\be
\mH_1 = \mH_A + \mH_\bA, \quad
\mH_A = {\rm span}\{|x_1\rangle,\;  x_1\in A\}, \;
\mH_\bA = {\rm span}\{|x_1\rangle,\;  x_1\in \bA \}
\label{ha-ha-bar}
\ee
To study the target space EE, we find it convenient to begin with a discussion of the two-fermion Hilbert space $\mH_2$ (we will come back to the one-particle case later on). The most general two-fermion wavefunction is
of the form
\footnote{Unspecified range of integration would mean full range.
E.g. $\int dx_1= \int_{\bf R} dx_1$.}
\bea
\kern-40pt &|\psi\rangle = \int dx_1 \int dx_2 \,
\psi(x_1,x_2)|x_1,x_2\rangle_a = \int dx_1 \int dx_2 \, \psi_a(x_1,x_2)|x_1,x_2\rangle ,\nonumber\\
\kern-70pt |x_1,x_2\rangle_a &\equiv \fr1{\sqrt{2!}}\left(| x_1\ran\otimes | x_2\ran- 
|x_2\ran \otimes| \lan x_1\ran \right), \;
\psi_a(x_1,x_2)\equiv \frac1{\sqrt{2!}}\left( \psi(x_1,x_2)-
\psi(x_2,x_1)\right)
\label{2-particle-psi}
\eea
The two-particle Hilbert space splits naturally into three sectors,
as follows:
\be
\mH_2 = \mH_{2,0} + \mH_{1,1} + \mH_{0,2}
\label{h2-decomp}
\ee
where
\begin{eqnarray}
\mH_{2,0} =& {\rm span}\{|x_1,x_2\rangle_a, \quad  x_1,x_2\in A \}\nonumber \\
\mH_{1,1}
=& {\rm span}\{|x_1,x_2\rangle_a,
\quad  x_1\in A, \,  x_2\in \bA\} \nonumber \\
\mH_{0,2} =& {\rm span}\{|x_1,x_2\rangle_a;\quad  x_1,x_2\in \bA\} \nonumber
\end{eqnarray}
In terms of the wavefunction \eq{2-particle-psi}, restricting ranges of the integrals over $x_1, x_2$ variously to the regions $A$, $\bA$ give the projection of the wavefunction to the various sectors: thus, e.g.
\be
|\psi\rangle_{1,1} = \int_A dx_1 \int_\bA dx_2 \,
\psi(x_1,x_2)|x_1,x_2\rangle_a,\;
_a\lan x_1,x_2 |\psi\ran = \psi_a(x_1,x_2)
\label{psi-20}
\ee
The corresponding projection operators $\Pi_{(p,q)} : \mH_2 \to \mH_{(p,q)}$ are
given by
\bea
& \Pi_{2,0} =\fr12 \int_{A,A} dx_1 dx_2 |x_1,x_2\ran_a \, _a\lan x_1,x_2 |
\nonumber\\
&\kern-50pt \Pi_{1,1} = \int_{A,\bA} dx_1 dx_2 |x_1,x_2\ran_a \, _a\lan x_1,x_2 | = \frac{1}{2} \left( \int_A dx_1 \int_{\bar{A}}dx_2 +\int_{\bar{A}} dx_1 \int_Adx_2 \right)|x_1,x_2\rangle_a\ {}_a\langle x_1,x_2|
\nonumber\\
& \Pi_{0,2} =\fr12 \int_{\bA,\bA} dx_1 dx_2 |x_1,x_2\ran_a \, _a\lan x_1,x_2| 
\eea
It is easy to see each of these projection operators squares to itself
and they add up to identity in $\mH_2$.

The generalization of these concepts to an $N$-fermion Hilbert space
is straightforward:
\be \mH_N = \oplus_{p,q; p+q=N} \mH_{p,q}
\label{h-n}
\ee Here the notation $\mH_{p,q}$ denotes a sector in which there are
$p$ particles in the region $A$ and $q$ particles in the complementary
region $\bA$.  We will denote by $\Pi_{p,q}$ ($p+q=N$) the projection
operators $\mH_N \to \mH_{p,q}$.

It is straightforward to generalize the above discussion to $N$
fermions in ${\bf R}^D$ and the target space region $A$ is defined by
a plane, say $A: \{x^1 > d, x^2,..,x^D \in {\bf R}\}$. $\bA= {\bf
	R}^D-A$.  E.g. if we denote the coordinates of the $N$ particles as
${\bf x}_i= x^I_i$, $i=1,2,...,N$, $I=1,2,...,D$, then the
wavefunctions belonging to $\mH_{p,q}$ are given by

\be
|\psi_{p,q} \ran = \int_A \prod_{I=1}^D \prod_{i=1}^p d^D{\bf x}_i\
\int_\bA \prod_{I=1}^D \prod_{i=p+1}^N d^D {\bf x}_i\ 
\psi(\{{\bf x}_i\}) |\{{\bf x}_i\}\ran_a
\label{psi-p-q-d}
\ee
where the subscript $a$ denotes antisymmetrization as before. The decomposition
\eq{h-n} is again true and the following discussion generalizes in a straightforward fashion with various one-dimensional integrals replaced by the corresponding $d$-dimensional integrals.

\gap2

\noindent{\bf Reduced density matrix (RDM)}

\gap2

We are interested in defining an RDM $\tilde\rho$,
associated with the region $A$, in a state $\rho$ in
the full Hilbert space (which could be pure or mixed). The RDM should
have the property that for observables $O$ which can be measured by
detectors in $A$, we should have, in an appropriate sense,
\be
\Tr(\rho O)=\Tr_A(\tilde \rho O) 
\label{def-RDM}
\ee
In the following we will define each side carefully.

In a QFT, when one is interested in a spatial subregion $A$ of space
time (as against target space), one proceeds by noting that the full
Hilbert space is a tensor product of the form $\mH= H_A \otimes
H_\bA$, which leads to $\tilde \rho= \Tr_{H_\bA} \rho$, with $\Tr_A$
interpreted as $\Tr_{H_A}$.

However, there is no such tensor product decomposition for target
space subregions. As we saw above, the single-particle Hilbert space
$\mH_1$ is a direct {\it sum}, rather than a product, of subspaces
associated with $A$ and $\bA$. A similar statement is true also for an
$N$-particle Hilbert space. What allows us to proceed is that each
given {\it sector} $\mH_{(p,q)}$ in an $N$-particle Hilbert space,
separately, has an (antisymmetric) tensor product of factors
associated with $A$ and $\bA$ respectively.

Let us explain the case of the $\mH_{1,1} \subset \mH_2$
as an illustration. It is easy to see that
\be
\mH_{1,1}= \mH_A \wedge \mH_\bA,\;
\label{tensor-11}
\ee
where the antisymmetric tensor product $V\wedge W$ denotes $V \otimes
W - W \otimes V$.

\noindent{\it Operator algebra}

The operators that map $\mH_{1,1} \to \mH_{1,1}$ are  of the form
\be
{\rm Span}\{ |x,y\ran_a\  _a\lan x',y'|,\ x,x'\in A,\ y,y' \in \bA\} 
\label{op-h11}
\ee
Among these, operators $O$ which correspond to observables in region $A$
must have the property
\be
O|x,y\ran_a = \int_A dx'\ \tilde O(x,x')|x',y\ran_a,
\label{op-alg-A-action}
\ee
which do not have any effect on $|y\ran$, $y\in \bA$. 
In fact, the corresponding operator algebra can 
be obtained by setting $y=y'$
in \eq{op-h11} and integrating over the $y$ coordinate. This gives
\footnote{The second
equality below can be derived as follows. Take a matrix element of
the operator inside the ``Span'' in the first line and show that it
is that in the second line.}

\bea {\cal A}_{1,1} &= {\rm
	Span}\{\int_\bA dy |x,y\ran_a\ _a\lan x',y|,\ x,x'\in A \}
\nonumber\\
&= {\rm Span}\{ |x\ran \lan x'|\otimes {\bf 1}_\bA + {\bf 1}_\bA \otimes |x\ran\ \lan x'|,\ x,x'\in A \}
\nonumber\\
{\bf 1}_\bA & \equiv \int_\bA\ dy\ | y\ran \lan y|
\label{op-alg-A}
\eea
It is easy to show that this operator algebra is closed under multiplication.
The operator $O$, with the action defined in \eq{op-alg-A-action}
can be identified as an element of \eq{op-alg-A}, with the form:
\be
\kern-10pt O= \tilde O_{1,1} \otimes {\bf 1}_\bA + {\bf
	1}_\bA \otimes \tilde O_{1,1}, \quad 
\tilde O_{1,1} \equiv \int_{A,A} dx dx' \
\tilde O(x, x')|x\ran \lan x'|,
\label{o11}
\ee
It is easy to check that this operator satisfies the defining property
\eq{op-alg-A-action} (note that ${\bf 1}_\bA |x\ran=0$ for $x\in A$).

\noindent{\it Density matrices}

A general state in the two-fermion Hilbert space is described by
a density matrix
\[
\rho = \int dx_1dx_2 \int dx'_1 dx'_2 \rho(x_1,x_2;x'_1, x'_2)
|x_1,x_2\ran_a\  _a\lan x_1',x_2'|
\]
The projection of $\rho$ onto the Hilbert space $\mH_{1,1}$ is given by
\be
\rho_{1,1}= \Pi_{1,1} \rho \Pi_{1,1}
= \int_{A,\bA} dx dy\int_{A,\bA} dx' dy'  \int_\bA dx_2 \rho(x,y;
x',y')|x,y\ran_a\  _a\lan x',y'|
\label{rho-11}
\ee
Although $\mH_{(1,1)}$ is not an usual tensor product but an
antisymmetrized one \eq{tensor-11}, one can define a partial trace
with respect to $\mH_\bA$ irrespective of the order of factors:
thus
\bea
\tilde \rho_{1,1} &=  \Tr_{\mH_\bA} \Pi_{1,1} \rho \Pi_{1,1}\nonumber\\
&= \int_\bA dy_1 \lan y_1| \ 
\left[\int_{A,\bA} dx dy\int_{A,\bA} dx' dy'  \int_\bA dx_2 \rho(x,y;
x',y')|x,y\ran_a\  _a\lan x',y'|\ \right] | y_1 \ran
\nonumber\\
& = \int_{A,A} dx dx' \int_\bA dy \rho(x,y;
x',y)|x\ran\ \lan x'|
\label{tilde-rho-11}
\eea
then we get
\[
\Tr_{\mH_2}\left(\rho\, O \right)
=\Tr_{\mH_A}\left(\tilde \rho_{1,1}\, \tilde O_{1,1} \right)
\]
which is of the form \eq{def-RDM}, except for the important
difference, characteristic of target space EE, that both operators on
the LHS are two-particle operators whereas those on the RHS are
one-particle operators defined on the one-particle Hilbert space
$\mH_A$ associated with the factor in \eq{tensor-11} associated with
region $A$; the traces on the two sides also pertain to these
two-particle and one-particle Hilbert spaces respectively.

In general, as mentioned above, an $N$-fermion Hilbert space $\mH_N$
splits into sectors $\mH_{p,q}$, $p+q=N$ (see \eq{h-n}). In each
$\mH_{p,q}$, there exists a tensor product decomposition into two
Hilbert spaces associated, respectively, with the regions $A$ and
$\bA$:
\be \mH_{p,q}= \mH_A^p \wedge \mH_\bA^q, \; \mH_A^p \equiv
\left(\wedge^p\mH_A \right),\, \mH_\bA^q \equiv \left(
\wedge^q\mH_\bA\right)
\label{gen-sector-decom}
\ee
where $\wedge^p V \equiv V \wedge V \wedge ...\wedge V$
($p$ times). For $V=\mH_A\ (\mH_\bA)$, respectively, these represent
$p$ fermions in region $A\ (\bA)$. By definition,
$\wedge^0 \mH_A =|0\ran_A= \mathcal{C}$ (zero particles in $A$)
and $\wedge^1 \mH_A=\mH_A$ (similarly for region $\bA$). Generalizing
\eq{tilde-rho-11}, it is easy to show that the RDM's in each sector
are given by
\be \tilde \rho_{p,q}= \Tr_{\mH^q_\bA} \rho_{p,q},\; \rho_{p,q} =\Pi_{p,q} \rho \Pi_{p,q}
\label{partial-trace-gen}
\ee
The target space EE is naturally given by the combined von Neumann entropy
of RDM's from all sectors:
\be
S= - \sum_{p,q; p+q=N} \Tr_{\mH_A^p} \tilde \rho_{p,q}\log(\tilde \rho_{p,q})
\label{target-space-EE}
\ee
This can be equivalently defined as
\[
S= - \Tr \tilde \rho\ \log(\tilde \rho)
\]
where $\tilde \rho$ is a formal sum of the sectorwise RDM's
\be
\tilde \rho = \oplus_{_{p+q=N}}\ \tilde \rho_{p,q}
\label{rho-tilde-app}
\ee
which acts on the sum of the vector spaces $\mH^p_A$ (in the notation of
\eq{gen-sector-decom}).\\

\noindent{\bf Explicit calculation ($N$=2)}\\

Let us work the EE in detail for the two-particle case
($N=2$). In this case, the various sectors have the tensor
decomposition
\be
\mH_{2,0}= \left( \wedge^2 \mH_A \right) \otimes \mathcal{C},\;
\mH_{1,1}= \mH_A \wedge \mH_\bA,\;
\mH_{2,0}= \mathcal{C} \otimes \left( \wedge^2 \mH_\bA \right)
\label{11-sectors-decom}
\ee
Note that a general 2 particle state $|\psi\ran$ can be written in multiple ways
\bea
|\psi\ran &=\int dx_1\int dx_2\ \psi(x_1,x_2) |x_1,x_2\ran_a \nonumber \\
&=\int dx_1\int dx_2\ \psi_a(x_1,x_2) |x_1,x_2\ran \nonumber \\
&=\frac1{\sqrt{2}}\int dx_1\int dx_2\ \psi_a(x_1,x_2) |x_1,x_2\ran_a
\eea
In the first line the kets are anti-symmetric, in the second line the wavefunction $\psi$ is anti-symmetric, while in the last line both the kets and the wavefunction $\psi$ are anti-symmetric. In the following we will use the last representation most often (as the symmetric part of the wavefunction $\psi$ never contributes if the kets are antisymmetrized). The density matrix $\rho$ is given  by
\begin{eqnarray}
\rho &=& |\psi\rangle\langle \psi| = \frac1{2}\int dx_1 dx_2 \int dx_1' dx_2'\ \psi_a(x_1,x_2)\psi_a^*(x_1',x_2') |x_1,x_2\ran_a\ {}_a\lan x_1',x_2'| \nonumber \\
&=& \frac1{2}\int dx_1 dx_2 \int dx_1' dx_2'\ \rho_a(x_1,x_2;x_1',x_2') |x_1,x_2\ran_a\ {}_a\lan x_1',x_2'|
\end{eqnarray}
To proceed, we follow \eq{partial-trace-gen} and \eq{11-sectors-decom}. In the (2,0) sector the partial trace over $\bA$ is trivial and we need to take care of only the projections, which just restrict the range of the integrals. Thus we get 
\bea
\tilde{\rho}_{2,0} &= {\rm Tr}_{\bar{A}}(\rho_{2,0})= \rho_{2,0} \nonumber\\
&=\frac1{2}\int_A dx_1 dx_2 \int_A dx_1' dx_2'\ \psi_a(x_1,x_2)\psi_a^*(x_1',x_2') |x_1,x_2\ran_a\ {}_a\lan x_1',x_2'|
\label{rho-2-0}
\eea
In $(1,1)$ sector, the projection operator (that restricts to $\mathcal{H}_{(1,1)}$) is given by $\Pi_{1,1} = \int_A dx_1 \int_{\bar{A}}dx_2 |x_1,x_2\rangle_a\ {}_a\langle x_1,x_2|$. First note its action on the ket $|\psi\rangle$
\begin{eqnarray}
\Pi_{1,1} |\psi\rangle &=& \frac1{\sqrt{2}}\int_A dx_1 \int_{\bar{A}}dx_2 |x_1,x_2\rangle_a {}_a\langle x_1,x_2| \int dy_1 dy_2\ \psi_a(y_1,y_2)|y_1 y_2\rangle_a \nonumber \\
&=& \frac1{\sqrt{2}}\int_A dx_1 \int_{\bar{A}}dx_2 |x_1,x_2\rangle_a \int dy_1 dy_2\ \psi_a(y_1,y_2) \times \nonumber \\
& &\left( \delta(x_1-y_1)\delta(x_2-y_2)-\delta(x_1-y_2)\delta(x_2-y_1)\right) \nonumber \\
&=& \sqrt{2}\int_A dx_1 \int_{\bar{A}}dx_2 \psi_a(x_1,x_2)|x_1,x_2\rangle_a \nonumber \\
&=& \frac1{\sqrt{2}} \left( \int_A dx_1 \int_{\bar{A}}dx_2 + \int_{\bar{A}} dx_1 \int_A dx_2\right) \psi_a(x_1,x_2)|x_1,x_2\rangle_a \nonumber
\end{eqnarray}
where we have used ${}_a\langle x_1,x_2| y_1 y_2\rangle_a = \delta(x_1-y_1)\delta(x_2-y_2)-\delta(x_1-y_2)\delta(x_2-y_1)$. Therefore
\begin{eqnarray}
\rho_{1,1} &=& 2\int_A dx_1 dx_1'\int_{\bar{A}} dx_2 dx_2'\ \psi_a(x_1,x_2) \psi_a^*(x_1',x_2')|x_1,x_2\rangle_a {}_a\langle x_1',x_2'| \nonumber
\end{eqnarray}
Now tracing over $\mathcal{H}_{\bar{A}}$
\begin{eqnarray}
\tilde{\rho}_{1,1} &= {\rm Tr}_{\mH_\bA} (\rho_{1,1}) = 2\int_A dx_1 dx_1'\int_{\bar{A}} dx_2 dx_2'\ \psi_a(x_1,x_2) \psi_a^*(x_1',x_2') \times \nonumber \\
&~~ \int_{\bar{A}}dz \langle z|\left(|x_1,x_2\rangle_a {}_a\langle x_1',x_2'|\right)|z\rangle \nonumber \\
&= 2\int_A dx_1 dx_1'\int_{\bar{A}} dx_2 dx_2'\ \psi_a(x_1,x_2) \psi_a^*(x_1',x_2') |x_1\rangle \langle x_1'| \langle x_2|x_2'\rangle \nonumber \\
&= 2\int_A dx_1 dx_1'\int_{\bar{A}} dx_2\ \psi_a(x_1,x_2) \psi_a^*(x_1',x_2) |x_1\rangle \langle x_1'| 
\label{rho-tilde-1-1}
\end{eqnarray}
In $(0,2)$ sector, $\Pi_{02} = \frac{1}{2}\int_{\bar{A}} dx_1 \int_{\bar{A}} dx_2 |x_1,x_2\rangle_a\ {}_a\langle x_1,x_2| $. After doing the appropriate
partial trace over $H(2,\bA)$ (see \eq{partial-trace-gen}), just gives
a number
\begin{equation}
 \tilde{\rho}_{0,2} = {\rm Tr}_{{\bar{A}}} (\rho_{0,2}) =\frac1{2}\int_{\bar{A}} dx_1 dx_2\, \psi_a(x_1,x_2)\psi_a^*(x_1,x_2)
\label{rho-0-2}
\end{equation}

As a specific example consider the Slater determinant state given by
\begin{equation}
|\psi\rangle 
= \frac{1}{\sqrt{2!}} \left(|u_1,u_2\rangle-|u_2,u_1\rangle \right) = \frac{1}{\sqrt{2!}}\sum_{i_1,i_2=1}^{2}\varepsilon_{i_1 i_2} |u_{i_1} u_{i_2}\rangle
\label{u1-u2}
\end{equation}
where $|u_i\ran= \int dx\ u_i(x) |x\ran $ are single particle wavefunctions. The wavefunction $\psi_a(x_1,x_2)=(u_1(x_1)u_2(x_2) -u_2(x_1)u_1(x_2) )/\sqrt{2}$. The corresponding density matrix is
\begin{eqnarray}
\rho = |\psi\rangle\langle \psi| &=&\frac{1}{2!} \left(|u_1,u_2\rangle-|u_2,u_1\rangle \right) \left(\langle u_1,u_2|-\langle u_2,u_1| \right) \nonumber \\
&=& \frac{1}{2!} \sum_{i_1,i_2,j_1,j_2=1}^{2} \varepsilon_{i_1 i_2} \varepsilon_{j_1 j_2} |u_{i_1} u_{i_2}\rangle \langle u_{j_1} u_{j_2}|
\label{rho-slater}
\end{eqnarray}
For this particular state the equations (\ref{rho-2-0}), (\ref{rho-tilde-1-1}) and (\ref{rho-0-2})) become
\begin{eqnarray}
\tilde{\rho}_{2,0} = \frac{1}{2!} \left(|u_1,u_2\rangle_{AA} -|u_2,u_1\rangle_{AA}\right) \left({}_{AA}\langle u_1,u_2|- {}_{AA}\langle u_2,u_1| \right) \label{rho20}\\
\kern-20pt \tilde{\rho}_{1,1} = \sum_{\rm i's,j's=1}^{2} \varepsilon_{i_1 i_2} \varepsilon_{j_1 j_2}  |u_{i_1}\rangle_A\  {}_A\langle u_{j_1}| \langle u_{j_2}|u_{i_2}\rangle_{\bar{A}} = 
\begin{blockarray}{l c c l}
& _A\langle u_1| & _A\langle u_2| & \\
\begin{block}{l [ c c ] l}
& (1-p_2) & -q_{21\bar{A}} & |u_1\rangle_A \\
& -q_{12\bar{A}} & (1-p_1) & |u_2\rangle_A \\
\end{block}
\end{blockarray} \label{rho11}\\
\tilde{\rho}_{0,2} =\frac{1}{2!} \sum_{i_1,i_2,j_1,j_2=1}^{2} \varepsilon_{i_1 i_2} \varepsilon_{j_1 j_2} \langle u_{j_1} u_{j_2}|u_{i_1} u_{i_2}\rangle_{\bar{A}} =(1-p_1)(1-p_2) - |q_{12\bar{A}}|^2 \label{rho02}
\end{eqnarray}
where $|u,v\ran_{AA}= |u\ran_A |v\ran_A$, $|u\ran_A= P_A |u\ran
\equiv \int_A dx\ u(x)|x\ran$, and $\lan u|v\ran_\bA \equiv \int_\bA dx\ u^*(x)v(x)$. Further we have written $\tilde{\rho}_{1,1}$ and $\tilde{\rho}_{0,2}$ in terms of $p_1=\int_A\ dx|u_1(x)|^2$, $p_2=\int_A\ dx|u_2(x)|^2$ and $q_{12\bar{A}}=\int_{\bar{A}}\ dx\ u_1^*(x)u_2(x)$.

Finally, following the general prescription \eq{target-space-EE},
the EE in target space is given by 
\[
S =- {\rm Tr} (\tilde{\rho} \log \tilde{\rho}) =
-\left[{\rm Tr} (\tilde{\rho}_{2,0} \log \tilde{\rho}_{2,0}) +{\rm Tr} (\tilde{\rho}_{1,1} \log \tilde{\rho}_{1,1}) +{\rm Tr} (\tilde{\rho}_{0,2} \log \tilde{\rho}_{0,2})\right]
\]

\gap2

\noindent{\bf Target space EE for general $N$}\\

Now we generalize this to a general N fermion state
\begin{equation}
|\psi\rangle = \frac{1}{\sqrt{N!}}\int dx_1\ldots x_N\ \psi_a(x_1,\ldots,x_N) |x_1\ldots x_N\rangle_a
\end{equation}
where we define
\begin{eqnarray}
\psi_a(x_1,\ldots,x_N) &\equiv& \frac{1}{\sqrt{N!}} \sum_{\sigma \in S_N} (-1)^\sigma \psi(x_{\sigma(1)},\ldots,x_{\sigma(N)})\\
|x_1\ldots x_N\rangle_a &\equiv& \frac{1}{\sqrt{N!}} \sum_{\sigma \in S_N} (-1)^\sigma |x_{\sigma(1)}\ldots x_{\sigma(N)}\rangle
\end{eqnarray}
The corresponding density matrix is
\begin{eqnarray}
&\kern-10pt \rho = |\psi\rangle \langle \psi| \nonumber \\
&\kern-10pt= \frac{1}{N!}\int dx_1 dx_1'\ldots x_N x_N'\ \psi_a(x_1,\ldots,x_N) \psi^*_a(x_1',\ldots,x_N') |x_1\ldots x_N\rangle_a\ {}_a\langle x_1'\ldots x_N'| \nonumber \\
&\kern-10pt= \frac{1}{N!}\int dx_1 dx_1'\ldots x_N x_N' \rho_a(x_1,\ldots,x_N;x_1',\ldots,x_N') |x_1\ldots x_N\rangle_a\ {}_a\langle x_1'\ldots x_N'|
\end{eqnarray}
For $N$ particle state we have $N+1$ sectors namely $(N,0)$, $(N-1,1)$,\ldots, $(0,N)$ where the first entry is the number of particles in $A$ and second entry in $\bar{A}$. In the $(k,N-k)$ sector, the projection operator is given by
\begin{equation}\label{eq:projection}
\kern-10pt \Pi_{k,N-k} = \frac{1}{N!} {N \choose k} \int_A dx_1 \ldots dx_k \int_{\bar{A}} dx_{k+1} \ldots dx_N |x_1,\ldots,x_N\rangle_a {}_a\langle x_1,\ldots, x_N|
\end{equation}
First note its action on $|\psi\rangle $
\begin{eqnarray}
&\Pi_{k,N-k}|\psi\rangle = \frac{1}{N!} {N \choose k} \int_A dx_1 \ldots dx_k \int_{\bar{A}} dx_{k+1} \ldots dx_N |x_1,\ldots,x_N\rangle_a \nonumber \\
& \times \frac{1}{\sqrt{N!}}\int dy_1\ldots y_N\ \psi_a(y_1,\ldots,y_N)\ {}_a\langle x_1,\ldots, x_N|y_1,\ldots,y_N\rangle_a \nonumber \\
&= \frac{1}{\sqrt{N!}}{N \choose k} \int_A dx_1 \ldots dx_k \int_{\bar{A}} dx_{k+1} \ldots dx_N\ \psi_a(x_1,\ldots,x_N) |x_1,\ldots,x_N\rangle_a
\end{eqnarray}
To go to the last line we have used
\begin{equation}
{}_a\langle x_1,\ldots, x_N|y_1,\ldots,y_N\rangle_a = \sum_{\sigma \in S_N} (-1)^\sigma \delta(x_1-y_{\sigma(1)})\ldots \delta(x_N-y_{\sigma(N)})
\end{equation}
The density matrix restricted to this sector is
\begin{eqnarray}
\kern-10pt \rho_{k,N-k} &=&\Pi_{k,N-k} \rho \Pi_{k,N-k} = \int_A dx_1 dx_1' \ldots dx_k dx_k' \int_{\bar{A}} dx_{k+1} dx_{k+1}'\ldots dx_N dx_N' \times \nonumber \\
& & \frac{1}{N!} {N \choose k}^2 \psi_a(x_1,\ldots,x_N) \psi_a(x_1',\ldots,x_N')^* |x_1,\ldots,x_N\rangle_a\ {}_a\langle x_1',\ldots,x_N'| \nonumber 
\end{eqnarray}
Next we need to trace over the particles in $\bar{A}$. This is easily done
\begin{eqnarray}\label{eq:sigma_general}
&\kern-10pt \tilde{\rho}_{k,N-k} = {\rm Tr}_{\bar{A}} (\rho_{k,N-k}) = \frac{1}{(N-k)!}\int_{\bar{A}} dz_{k+1}\ldots dz_N\ {}_a\langle z_{k+1}\ldots z_N| \rho_{k,N-k} |z_{k+1}\ldots z_N \rangle_a \nonumber \\
&\kern-10pt =\frac{1}{N!} {N \choose k}^2 \int_A dx_1 dx_1' \ldots dx_k dx_k' \int_{\bar{A}} dx_{k+1} dx_{k+1}'\ldots dx_N dx_N'\ \psi_a(x_1,\ldots,x_N) \times \nonumber \\
&~ \psi^*_a(x_1',\ldots,x_N') |x_1,\ldots, x_k\rangle_a {}_a\langle x_1'\ldots, x_k'|\ {}_a\langle x_{k+1},\ldots,x_N|y_{k+1}\ldots,y_N\rangle_a \nonumber \\
&\kern-10pt =\frac{1}{k!} {N \choose k} \int_A dx_1 dy_1 \ldots dx_k dy_k \int_{\bar{A}} dx_{k+1} \ldots dx_N\ \psi_a(x_1,\ldots,x_N) \psi^*_a(x_1',\ldots,x_N') \times \nonumber \\
&~ |x_1,\ldots, x_k\rangle_a\ {}_a\langle x_1'\ldots, x_k'|
\end{eqnarray}
where we have used
\begin{eqnarray}
&\kern-10pt \frac{1}{(N-k)!} \int_{\bar{A}} dz_{k+1}\ldots dz_N\ {}_a\langle z_{k+1}\ldots z_N|x_{k+1},\ldots,x_N\rangle_a {}_a\langle x_{k+1}',\ldots,x_N'|z_{k+1}\ldots z_N \rangle_a \nonumber \\
&= |x_1\ldots x_k\rangle_a\ {}_a\langle x_1\ldots x_k|\ {}_a\langle y_{k+1}\ldots y_N|x_{k+1},\ldots,x_N\rangle_a
\end{eqnarray} 

More specifically consider the state given by a Slater determinant (of single-particle states $u_1,u_2,\ldots,u_N $)
\begin{equation}\label{eq:slater_state}
|\psi\rangle = \frac{1}{\sqrt{N!}}\sum_{i's} \varepsilon_{i_1 \ldots i_N} |u_{i_1}\ldots u_{i_N}\rangle
\end{equation}
with the wavefunction $\psi_a(x_1,\ldots,x_N)=\sum_{i\rm{'s}} \frac{1}{\sqrt{N!}} \varepsilon_{i_1 \ldots i_N} u_{i_1}(x_1)\ldots u_{i_N}(x_N) $. Each of the $i$'s can take values from 1 to N, i.e. $i_n \in \{1,2,\ldots,N\}$. The RDM in the $(k,N-k)$ sector (\ref{eq:sigma_general}) is given by
\begin{equation}\label{eq:sigma_slater}
\kern-20pt \tilde{\rho}_{k,N-k} = {N \choose k} \frac{1}{N!}\sum_{i\rm{'s}, j\rm{'s}} \varepsilon_{i_1 \ldots i_N} \varepsilon_{j_1 \ldots j_N} |u_{i_1} \ldots u_{i_k}\rangle_A\ {}_A\langle u_{j_1} \ldots u_{j_k}| \prod_{n=k+1}^{N}\langle u_{j_n}| u_{i_n}\rangle_{\bar{A}}
\end{equation}
This formula is very simple to understand. If we worked with the position space wavefunctions $\langle x_1\ldots x_N|\psi\rangle= \psi_a(x_1,\ldots,x_N)$, the RDM is simply given by
\begin{eqnarray}
\kern-20pt \tilde{\rho}(x1,\ldots,x_k;x1',\ldots,x_k') &=& {N \choose k}\int_{\bar{A}} dx_{k+1}\ldots dx_N \psi_a(x_1,\ldots,x_k,x_{k+1}\ldots x_N) \times \nonumber \\
& & \psi^*_a(x_1',\ldots,x_k',x_{k+1}\ldots x_N)
\end{eqnarray}
with the factor ${N \choose k}$ coming from the number of ways choosing the integration variables. This is the origin of ${N \choose k}$ in (\ref{eq:sigma_general}) and (\ref{eq:sigma_slater}) while the remaining numerical factor is just for normalization.

The EE is by the general formula given above \eq{target-space-EE}:
\begin{equation}\label{eq:tsee}
S =- {\rm Tr} \tilde{\rho} \log \tilde{\rho} = -\sum_k {\rm Tr}\, \tilde{\rho}_{k,N-k} \log \tilde{\rho}_{k,N-k}
\end{equation}

\paragraph{\textbf {Equivalence of 1st and 2nd quantized entanglement entropy for free theories}}

\subsection{2nd quantized theory}

The target space subregion $A \subset R$ in the first quantized
formalism, can be viewed as a spatial subregion from the viewpoint of
the second quantized formalism where the single particle states $|x\ran$
can be regarded as created from the zero particle state $|0\ran$ by
the second quantized creation operator:
\[
|x\ran= \Psi^\dagger(x)|0\ran
\]
The general Fock space state 
can be regarded as a linear combination of the antisymmetric
states
\[
{\cal F} \ni |x_1, x_2,..., x_N\ran_a =\frac1{\sqrt{N!}}\sum_{\sigma\in S(N)}
| x_{\sigma(1)},..., x_{\sigma(N)}\ran
=
\Psi^\dagger(x_1)
... \Psi^\dagger(x_N)|0\ran
\]
It is easy to see the tensor product decomposition
\be
{\cal F}= {\cal F}_A \wedge {\cal F}_\bA
\label{fock-tensor}
\ee
which allows one to define RDM's in terms of the usual partial traces.

Note that since each Fock space is a sum of 0,1,2,.. particle Hilbert
spaces, we can write, using the notations in \eq{gen-sector-decom}:
\[
{\cal F}_A = |0\ran_A \oplus \mH_A  \oplus \mH_A^2 \oplus ..., \quad
{\cal F}_\bA = |0\ran_\bA \oplus \mH_\bA  \oplus \mH_\bA^2 \oplus ...
\]
The tensor product \eq{fock-tensor} thus gets written as a direct sum
\bea
&{\cal F}= \left(|0\ran\right) \oplus \left(\mH_A  \oplus \mH_\bA\right)
\oplus \left(\mH_A^2  \oplus (\mH_A \wedge \mH_\bA)
\oplus \mH_\bA^2  \right) + ... \nonumber\\
&= \mH_0 \oplus \mH_1 \oplus \mH_2 + ...
\label{fock-detail}
\eea
Here $|0\ran_A \equiv \mH_A^0 \equiv C$ is the zero-particle state in $A$,
defined by $\Psi(x)|0\ran_A=0$ for all $x\in A$ (similarly for $\bA$);
we have, further used the identities:
\[
|0\ran_A \otimes |0\ran_\bA = |0\ran,\;  |0\ran_A \otimes \mH_\bA^p
= \mH_\bA^p,\;  \mH_\bA^p \otimes |0\ran_\bA
= \mH_\bA^p,
\]
Note that a wedge product with zero-particle states such as $|0\ran_A$
becomes an ordinary tensor product (it amounts to just scalar
multiplication by a complex number, see below \eq{gen-sector-decom}).

Written in the form \eq{fock-detail}, we can clearly identify the
terms in round brackets as the first quantized Hilbert spaces $\mH_n$
with a clear sum of products structure introduced in \eq{h-n},
\eq{gen-sector-decom}. We will find below that the RDM in the second
quantized framework, sector by sector, is the same as that in the
first quantized framework.\\

\noindent{\bf Computation of RDM}\\

EE in field theories is well-studied in the literature and we follow the method of \cite{chuerta}. Using the decomposition \eq{fock-tensor}, the reduced density matrix $\rho_A$ of a region $A$, is defined by
\[
\rho_A= \Tr_{{\cal F}_A}\rho
\]
where $\rho$ is the density matrix corresponding to the state of the full system. This is, of course, an operator in ${\cal F}_A$; however, as shown in \cite{chuerta}, 
for theories with quadratic modular Hamiltonian as in the case for free fermions, $\rho_A$ can be expressed in terms of the exponential of a one-body (particle-number preserving) operator $\hat H_A$, the so-called modular hamiltonian:
\be
\rho_A = K e^{-\hat H_A}.
\label{mod-ham}
\ee
where $K$ is a constant ensuring $\Tr_{{\cal F}_A}\rho_A=1$. 
The modular hamiltonian, projected onto the one-particle Hilbert space $\mH_A$,
(let us call it $\hat H_{A}^{(1)}$) can be expressed in an orthonormal basis of $\mH_A$:
\[
\hat H_{A}^{(1)}= \sum_l \epsilon_l |l\ran \lan l|
\]
By definition, $\lan l | l'\ran= \delta_{ll'}$ and $v_l(x)
\equiv \lan x|l\ran$ has support only in $x\in A$. 
Defining creation and annihilation operators $d_l, d^\dagger_l$
such that $|l\ran = d^\dagger |0\ran$, clearly the Fock space operator will be given by
$\hat H_A = \sum_l \epsilon_l d_l^\dagger d_l$. Using this and \eq{mod-ham},
we get
\[
\rho_A= K \exp\left[- \sum_l \epsilon_l d_l^\dagger d_l \right] =
\prod_{l} \frac{e^{- \epsilon_l d_l^\dagger d_l}}{(1+e^{-\epsilon_l})}
\]
Suppose we restrict to the N-particle sector of the full Folk space ${\cal F}$;
by \eq{fock-detail} this sector will have contributions from $\mH_k,\;
k=0,1,...,N$. We find that to describe this situation it is enough to keep only the first N number of $\l_i$'s non-zero (the corresponding $\epsilon_i$s finite) while all other $\l_i$'s can be set equal to zero (the corresponding $\epsilon_i$'s sent to infinity). Therefore one can write the N-particle density matrix as
\begin{equation}
\rho_A^{(N)} = \prod_{l=1}^N \frac{e^{- \epsilon_l d_l^\dagger d_l}}{(1+e^{-\epsilon_l})}
\label{rho-A-N}
\end{equation}

\noindent{\it Two particles ($N=2$)}

First we explicitly work out the 2 particle case and then generalize to arbitrary $N$. The density matrix for $N=2$ is
\be \rho_A^{(2)}= \frac{e^{- \epsilon_1 d_1^\dagger d_1}}{1+e^{-\epsilon_1}} \frac{e^{- \epsilon_2 d_2^\dagger d_2}}{1+e^{-\epsilon_2}}
\label{rho-A-2}
\ee
Without loss of generality, consider the following two particle state in the full space
\[ |s\rangle=b_2^\dagger b_1^\dagger|0\rangle \]
where $b, b^\dagger$'s are `global' fermionic annihilation/creation operators satisfying the standard algebra $\{b_i,b_j^\dagger\} =\delta_{ij}$. The one-particle states $|i\ran= b^\dagger_i |0\ran$ are global states, i.e. $u_i(x)
\equiv \lan x|i\ran$ have support in $x\in {\bf R}= A \cup \bA$.

The second quantized field $\Psi(x)$ has mode expansions of the form
\bea \Psi(x) &=\sum_i u_i(x) b_i, \; x\in {\bf R},\; u_i(x)= \lan
x| b^\dagger_i |0\ran, \; \int_{\bf R} dx u^*_i(x)u_j(x) =\delta_{ij}
\nonumber\\ \Psi(x) &=\sum_l v_l(x) d_l, \; x\in A,\; v_l(x)= \lan
x| d^\dagger_l |0\ran, \; \int_{A} dx v^*_l(x)v_m(x) =\delta_{lm}
\label{mode-exp}
\eea
The corresponding formulae for $\Psi^\dagger(x)$ are given by taking hermitian conjugation of the above equations. 
If $\rho_A^{(2)}$ is indeed the correct density matrix for the region of interest, the following equations should be true (as long as all the
operator insertions are within region $A$)
\begin{eqnarray}\label{eq:2particle-identities}
{\rm Tr} \left(\rho_A^{(2)} \Psi^\dagger(x_1) \Psi^\dagger(x_2) \Psi(x_1') \Psi(x_2') \right) &=& \langle s|\Psi^\dagger(x_1) \Psi^\dagger(x_2) \Psi(x_1') \Psi(x_2') |s\rangle \nonumber \\
{\rm Tr} \left(\rho_A^{(2)} \Psi^\dagger(x_1) \Psi(x_1') \right) &=& \langle s|\Psi^\dagger(x_1) \Psi(x_1') |s\rangle \nonumber \\
{\rm Tr} \left(\rho_A^{(2)} \right) &=& \langle s|s\rangle
\end{eqnarray}
Using the mode expansions \eq{mode-exp}, the equations (\ref{eq:2particle-identities}) lead to (respectively) 
\begin{eqnarray}\label{eq:2particle-identities-2}
\lambda_1 \lambda_2 \begin{array}{|cc|}
v_1(x_1) & v_2(x_1) \\
v_1(x_2) & v_2(x_2)
\end{array}^*
\begin{array}{|cc|}
v_1(x_1') & v_2(x_1') \\
v_1(x_2') & v_2(x_2')
\end{array}
&=& \begin{array}{|cc|}
u_1(x_1) & u_2(x_1) \\
u_1(x_2) & u_2(x_2)
\end{array}^*
\begin{array}{|cc|}
u_1(x_1') & u_2(x_1') \\
u_1(x_2') & u_2(x_2')
\end{array} \nonumber \\
\lambda_1 v_1(x_1)^* v_1(x_1') + \lambda_2 v_2(x_1)^* v_2(x_1') &=& u_1(x_1)^* u_1(x_1') + u_2(x_1)^* u_2(x_1')\nonumber \\
{\rm Tr} \left(\rho_A^{(2)} \right) &=& 1
\end{eqnarray}
where $\lambda_i = e^{-\epsilon_i}/(1+e^{-\epsilon_i})$. The last equation above just says that our density matrix should be properly normalized. The remaining two can be written more compactly as operator equations and through the use of generalized Kronecker delta functions as
\begin{eqnarray}
\sum_{i{\rm 's},j{\rm 's}=1}^2 \lambda_{i_1}\lambda_{i_2} \delta_{i_1 i_2}^{j_1 j_2} |v_{i_1}\rangle |v_{i_2}\rangle \langle v_{j_1}| \langle v_{j_2}| &=& \sum_{i{\rm 's},j{\rm 's}=1}^2 \delta_{i_1 i_2}^{j_1 j_2} |u_{i_1}\rangle_A |u_{i_2}\rangle_A~{}_A\langle u_{j_1}|{}_A\langle u_{j_2}| \nonumber \\
\sum_{i_1,j_1=1}^2 \lambda_{i_1} \delta_{i_1}^{j_1} |v_{i_1}\rangle \langle v_{j_1}| &=& \sum_{i_1,j_1=1}^2 \delta_{i_1}^{j_1} |u_{i_2}\rangle_A~{}_A\langle v_{j_1}|
\end{eqnarray}
where $|u_i\rangle_A = \int_A dx~u_i(x)|x\rangle $. This follows since the the relation (\ref{eq:2particle-identities-2}) is true for all $x_i,x_j' \in A$, we can multiply by position kets and integrate over region $A$. The generalized Kronecker delta function $\delta_{i_1 \ldots i_n}^{j_1 \ldots j_n}$ is defined to be +1(-1) when $i_1 \ldots i_n$'s are distinct and even(odd) permutation of $j_1 \ldots j_n$'s, otherwise it is 0.

From the structure of \eq{rho-A-2}, it is clear that $\rho_A^{(2)}$ has
non-zero matrix elements only in the four-dimensional Hilbert space spanned by\\
(a) $d_2^\dagger d_1^\dagger|0\rangle$,\\
(b) $d_1^\dagger|0\ran,\, d_2^\dagger|0\rangle$ and \\ 
(c) $|0\ran$,\\
representing, respectively, a two-particle state, two one-particle states and the zero-particle state in $\mH_A$. It is easy to see that these states
provide an eigenbasis of \eq{rho-A-2} with eigenvalues\\
(a) $\l_1\l_2$,\\
(b) $\l_1(1-\l_2), \l_2(1-\l_2)$, and\\
(c) $(1-\l_1)(1-\l_2)$,\\
respectively.

Using these facts, we can write the density matrix restricted to
the two-particle subsector (a), as follows
\[
\rho_{A,2}^{(2)}= \lambda_1 \lambda_2 d_2^\dagger d_1^\dagger|0\rangle \langle 0|d_1 d_2 = \lambda_1 \lambda_2 \frac{1}{\sqrt{2!}}(|v_1 v_2\rangle-|v_2 v_1\rangle) \frac{1}{\sqrt{2!}}(\langle v_1 v_2|-\langle v_2 v_1| )
\]
Using the first identity in (\ref{eq:2particle-identities-2}) one can write
\[\rho_{A,2}^{(2)}= \frac{1}{\sqrt{2!}}(|u_1 u_2\rangle_A-|u_2 u_1\rangle_A) \frac{1}{\sqrt{2!}}({}_A\langle u_1 u_2|-{}_A\langle u_2 u_1| )\]
The subscript $A$ is there to remind that this operator has support only in region $A$. Notice that this precisely matches the first quantized density matrix $\tilde \rho_{2,0}$ \eq{rho20}.

Using the eigenvalues mentioned above, the density matrix, restricted
to the one-particle subsector (b), can be written as  a diagonal matrix in the following basis
\[
\begin{blockarray}{l c c l}
& d_1^\dagger|0\rangle & d_2^\dagger|0\rangle & \\
\begin{block}{l [ c c ] l}
\rho_{A,1}^{(2)}= & \lambda_1 (1-\lambda_2) & 0 &~ \langle 0|d_1 \\
& 0  & (1-\lambda_1) \lambda_2 & ~\langle 0|d_2 \\
\end{block}
\end{blockarray}
\]
Through successive use of the identities (\ref{eq:2particle-identities-2}) we can write
\begin{eqnarray}
&\kern-10pt \rho_{A,1}^{(2)}= \lambda_1 (1-\lambda_2)|v_1\rangle\langle v_1| + \lambda_2 (1-\lambda_1)|v_2\rangle\langle v_2| \nonumber \\
&\kern-10pt = -\lambda_1\lambda_2 \left(|v_1\rangle\langle v_1| + |v_2\rangle\langle v_2|\right) + \left( \lambda_1 |v_1\rangle\langle v_1| + \lambda_2 |v_2\rangle\langle v_2| \right) \nonumber \\
&\kern-10pt = -\lambda_1\lambda_2 \left(|v_1\rangle\langle v_1| \langle v_2|v_2\rangle  + |v_2\rangle\langle v_2| \langle v_1|v_1\rangle - |v_1\rangle\langle v_2| \langle v_1|v_2\rangle - |v_2\rangle\langle v_1| \langle v_2|v_1\rangle \right) \nonumber \\
&~ + \left( |u_1\rangle_A {}_A\langle u_1| + |u_2\rangle_A {}_A\langle u_2| \right) \nonumber \\
&\kern-10pt = -(|u_1\rangle_A {}_A\langle u_1| {}_A\langle u_2|u_2\rangle_A  + |u_2\rangle_A {}_A\langle v_2| {}_A\langle u_1|u_1\rangle_A - |u_1\rangle_A {}_A\langle u_2| {}_A\langle u_1|u_2\rangle_A \nonumber \\
&~ -|u_2\rangle_A {}_A\langle u_1| {}_A\langle u_2|u_1\rangle_A ) +\left( |u_1\rangle_A {}_A\langle u_1| + |u_2\rangle_A {}_A\langle u_2| \right)
\end{eqnarray}
where in going from 2nd to 3rd line we have used the 2nd equation in (\ref{eq:2particle-identities-2}) for the second term. In the 3rd line, we have also introduced inner products of $v$'s (which are orthonormal) so that we could use the 1st equation in (\ref{eq:2particle-identities-2}) leading to the final expression. This matches precisely with the 1st quantized density matrix $\tilde \rho_{1,1}$ (\ref{rho11}).

The density matrix $\rho_A^{(2)}$, restricted to the zero-particle subsector (c) (let us
call it $\rho_{A,0}^{(2)}$) is proportional to $|0\ran \lan 0|$, and agrees with the corresponding first quantized quantity $\tilde \rho_{0,2}$ (this can be directly verified from the two-particle identities \eq{eq:2particle-identities-2}).

Thus we see that in our 2 particle example, the density matrices $\rho_{A,2}^{(2)},\rho_{A,1}^{(2)},\rho_{A,0}^{(2)}$ match with
$\tilde{\rho}_{2,0},\tilde{\rho}_{1,1},\tilde{\rho}_{0,2}$ (respectively) in the first quantized language.\\

\noindent{\it Arbitrary N}\\

Now we move to {\it arbitrary $N$}. Similar to the 2 particle example we employ the use of following identities (for $0\leq n\leq N$)
\begin{equation}
\kern-40pt {\rm Tr} (\rho_A^{(N)} \Psi^\dagger(x_1) \ldots \Psi^\dagger(x_n) \Psi(x_1') \ldots \Psi(x_n')) = \langle s|\Psi^\dagger(x_1) \ldots \Psi^\dagger(x_n) \Psi(x_1') \ldots \Psi(x_n') |s\rangle
\end{equation}
which is true as long as all the insertions $x_1,\ldots ,x_{n}'$ lie in region A. The state $|s\rangle = b_1^\dagger\ldots b_N^\dagger|0\rangle $ is a global $N$-particle state in the full space. Using the appropriate mode expansion for $\Psi(x)$ and after a bit of algebra, these identities can be written as
\begin{eqnarray}
&\sum_{i\rm{'s}, j\rm{'s}} \delta_{i_1 \ldots i_{n}}^{j_1 \ldots j_{n}} \lambda_{i_1}\ldots \lambda_{i_n} v_{j_1}^*(x_1) \ldots v_{j_n}^*(x_n) v_{i_1}(x_1') \ldots v_{i_n}(x_n') \nonumber \\
&= \sum_{i\rm{'s}, j\rm{'s}} \delta_{i_1 \ldots i_{n}}^{j_1 \ldots j_{n}} u_{j_1}^*(x_1) \ldots u_{j_n}^*(x_n) u_{i_1}(x_1') \ldots u_{i_n}(x_n')
\end{eqnarray}
where
\[
\lambda_i = \frac{e^{-\epsilon_i}}{1+e^{-\epsilon_i}}
\]
We can also write the identities as an operator equation
\begin{equation}\label{eq:identities}
\kern-30pt \sum_{i\rm{'s}, j\rm{'s}} \delta_{i_1 \ldots i_{n}}^{j_1 \ldots j_{n}} \lambda_{i_1}\ldots \lambda_{i_n} |v_{i_1} \ldots v_{i_n}\rangle \langle v_{j_1} \ldots v_{j_n}| = \sum_{i\rm{'s}, j\rm{'s}} \delta_{i_1 \ldots i_{n}}^{j_1 \ldots j_{n}} |u_{i_1} \ldots u_{i_n} \rangle_A\ {}_A\langle u_{j_1} \ldots u_{j_n}|
\end{equation}
Further when there only $k$ particles in region $A$ (out of $N$), one can easily write the $\rho^{(N)}_k$ in region $A$ as ${N\choose k} \times {N\choose k}$ diagonal matrix in the following basis
\begin{equation}
\begin{blockarray}{l c c c l}
& d_1^\dagger \ldots d_k^\dagger|0\rangle & \ldots & d_{N-k+1}^\dagger \ldots d_N^\dagger|0\rangle & \\
\begin{block}{l [ c c c ] l}
& \prod_{i=1}^{k}\lambda_i \prod_{j=k+1}^{N}(1-\lambda_{j}) & \ldots & 0 & \langle 0|d_k\ldots d_1 \\
\kern-40pt \rho_{A,k}^{(N)}= &0  &  \ldots & 0 & \ldots \\
& \ldots & \ldots & \ldots & \ldots \\
& 0 & \ldots & \prod_{i=1}^{N-k}(1-\lambda_{i}) \prod_{j=N-k+1}^{N}\lambda_j & \langle 0|d_N\ldots d_{N-k+1} \\
\end{block}
\end{blockarray}
\end{equation}
For example consider $N=3$ and $k=2$
\begin{equation}
\begin{blockarray}{l c c c l}
& d_1^\dagger d_2^\dagger|0\rangle & d_2^\dagger d_3^\dagger|0\rangle & d_3^\dagger d_1^\dagger|0\rangle & \\
\begin{block}{l [ c c c ] l}
& \lambda_1 \lambda_2 (1-\lambda_3) & 0 & 0 & \langle 0|d_2 d_1 \\
\rho_{A,2}^{(3)}= & 0  & (1-\lambda_1) \lambda_2 \lambda_3 & 0 & \langle 0|d_3 d_2 \\
& 0 & 0 & \lambda_1 (1-\lambda_2) \lambda_3 & \langle 0|d_1 d_3 \\
\end{block}
\end{blockarray}
\end{equation}
which is a ${3\choose 2} \times {3\choose 2}$ matrix. We can write $\rho^{(N)}_k$ in a more compact notation
\begin{equation}
\kern-10pt \rho^{(N)}_{A,k} = \frac{1}{k!}\sum_{i\rm{'s}, j\rm{'s}} \delta_{i_1 \ldots i_{k}}^{j_1 \ldots j_{k}} |v_{i_1} \ldots v_{i_k}\rangle \langle v_{j_1} \ldots v_{j_k}| \lambda_{i_1}\ldots \lambda_{i_k} \prod_{m=1,m\neq i_1, \ldots, m\neq i_k}^N (1-\lambda_m)
\end{equation}
We can rewrite the generalized Kronecker delta function in terms of the Levi-Civita symbols using the following identity
\begin{equation}\label{eq:levicivita-kronecker}
\sum_{i\rm{'s}, j\rm{'s}} \varepsilon_{i_1 \ldots i_N} \varepsilon_{j_1 \ldots j_N} \delta_{j_{k+1} i_{k+1}} \ldots \delta_{j_{N} i_{N}} = (N-k)!\, \delta_{i_1 \ldots i_k}^{j_1 \ldots j_k}
\end{equation}
Making use of this we write
\begin{eqnarray}
\rho^{(N)}_{A,k}&= \frac{1}{k!(N-k)!}\sum_{i\rm{'s}, j\rm{'s}} \varepsilon_{i_1 \ldots i_N} \varepsilon_{j_1 \ldots j_N} |v_{i_1} \ldots v_{i_k}\rangle \langle v_{j_1} \ldots v_{j_k}|~ \lambda_{i_1}\ldots \lambda_{i_k} \times \nonumber \\
&~~~ \delta_{j_{k+1} i_{k+1}} \ldots \delta_{j_N i_N} \prod_{m=1,m\neq i_1, \ldots, m\neq i_k}^N (1-\lambda_m) \nonumber \\
& = \frac{1}{k!(N-k)!}\sum_{i\rm{'s}, j\rm{'s}} \varepsilon_{i_1 \ldots i_N} \varepsilon_{j_1 \ldots j_N} |v_{i_1} \ldots v_{i_k}\rangle \langle v_{j_1} \ldots v_{j_k}|~ \lambda_{i_1}\ldots \lambda_{i_k} \times \nonumber \\
&~~~ (1-\lambda_{i_{k+1}})\ldots (1-\lambda_{i_N}) \delta_{j_{k+1} i_{k+1}} \ldots \delta_{j_N i_N} \nonumber \\
& = \frac{1}{k!(N-k)!}\sum_{i\rm{'s}, j\rm{'s}} \varepsilon_{i_1 \ldots i_N} \varepsilon_{j_1 \ldots j_N} |v_{i_1} \ldots v_{i_k}\rangle \langle v_{j_1} \ldots v_{j_k}|~ \delta_{j_{k+1} i_{k+1}} \ldots \delta_{j_N i_N} \times \nonumber \\
&~~~ \left\{\sum_{l=0}^{N-k} (-1)^l {N-k \choose l} \lambda_{i_1}\ldots \lambda_{i_{k+l}} \right\} \nonumber \\
\end{eqnarray}
where in the last line we have opened the product over ($1-\lambda$)'s and organized the sum in powers of $\lambda_i$'s. In a particular $l$ term in the sum, since the $v_i$'s are orthonormal in region $A$, we can replace $\delta_{j_{k+1} i_{k+1}}\ldots \delta_{j_{k+1} i_{k+1}}$ by $\langle v_{j_{k+1}}|v_{i_{k+1}}\rangle_A \ldots \langle v_{j_{k+l}}|v_{i_{k+l}}\rangle_A $ but leave the remaining $\delta_{j_{k+l+1} i_{k+l+1}}\ldots \delta_{j_N i_N}$ as it is. After this we use (\ref{eq:levicivita-kronecker}) again with the remaining delta functions to go back to the generalized Kronecker delta $\delta_{i_1 \ldots i_{k+l}}^{j_1 \ldots j_{k+l}}$, this leads to
\begin{eqnarray}
&\kern-70pt \rho^{(N)}_{A,k}= \frac{1}{k!} \sum_{l=0}^{N-k} \frac{(-1)^l}{l!} \sum_{i\rm{'s}, j\rm{'s}} \delta_{i_1 \ldots i_{k+l}}^{j_1 \ldots j_{k+l}} |v_{i_1} \ldots v_{i_k}\rangle \langle v_{j_1} \ldots v_{j_k}| \lambda_{i_1}\ldots \lambda_{i_{k+l}} \langle v_{j_{k+1}}|v_{i_{k+1}}\rangle_A \ldots \langle v_{j_{k+l}}|v_{i_{k+l}}\rangle_A \nonumber \\
&\kern-70pt = \frac{1}{k!} \sum_{l=0}^{N-k} \frac{(-1)^l}{l!} \Bigg[ \sum_{i\rm{'s}, j\rm{'s}} \delta_{i_1 \ldots i_{k+l}}^{j_1 \ldots j_{k+l}} |u_{i_1} \ldots u_{i_k}\rangle_A~ {}_A\langle u_{j_1} \ldots u_{j_k}| \langle u_{j_{k+1}}|u_{i_{k+1}}\rangle_A \ldots \langle u_{j_{k+l}}|u_{i_{k+l}}\rangle_A \Bigg] \label{eq:rho-2}
\end{eqnarray}
In the last step we used the identities (\ref{eq:identities}).

\subsection{Comparison with the first Quantized Theory}

We will now show that the density matrix within any given sector (\ref{eq:sigma_slater}) agrees with its counterpart \eq{eq:rho-2}.

To begin, notice that the density matrix within a sector (\ref{eq:sigma_slater}) has inner products in region $\bar{A}$. We would like to write it in terms of region $A$ since that is what naturally appears in the second quantized theory. Using the orthonormality of $u_n(x)$'s
\[ \langle u_{j_{n}}| u_{i_{n}}\rangle_{\bar{A}} = \delta_{j_{n} i_{n}} -\langle u_{j_{n}}| u_{i_{n}}\rangle_{A} \]
We plug this in (\ref{eq:sigma_slater})
\begin{eqnarray}
&\kern-40pt \tilde{\rho}_{k,N-k} = \frac{{N \choose k}}{N!} \sum_{i\rm{'s}, j\rm{'s}} \varepsilon_{u_{i_1} \ldots u_{i_N}} \varepsilon_{u_{j_1} \ldots u_{j_N}} |u_{i_1} \ldots u_{i_k}\rangle_A\ {}_A\langle u_{j_1} \ldots u_{j_k}| \prod_{n=k+1}^{N}\langle u_{j_{n}}| u_{i_{n}}\rangle_{\bar{A}} \nonumber \\
&\kern-40pt = \frac{{N \choose k}}{N!} \sum_{i\rm{'s}, j\rm{'s}} \varepsilon_{i_1 \ldots i_N} \varepsilon_{j_1 \ldots j_N} |u_{i_1} \ldots u_{i_k}\rangle_A\ {}_A\langle u_{j_1} \ldots u_{j_k}| \prod_{n=k+1}^{N} \left( \delta_{j_{n} i_{n}} -\langle u_{j_{n}}| u_{i_{n}}\rangle_{A} \right) \nonumber \\
&\kern-40pt = \frac{{N \choose k}}{N!} \sum_{i\rm{'s}, j\rm{'s}} \varepsilon_{i_1 \ldots i_N} \varepsilon_{j_1 \ldots j_N} |u_{i_1} \ldots u_{i_k}\rangle_A\ {}_A\langle u_{j_1} \ldots u_{j_k}| \Bigg[\prod_{n=k+1}^{N} \delta_{j_{n} i_{n}} - (N-1) \langle u_{j_{k+1}}|u_{i_{k+1}}\rangle_A \nonumber \\
&\kern-30pt \prod_{n=k+2}^{N} \delta_{j_{n} i_{n}} + \ldots + (-1)^l {N-k \choose l} \langle u_{j_{k+1}}|u_{i_{k+1}}\rangle_A \langle u_{j_{k+2}}|u_{i_{k+2}}\rangle_A \ldots \langle u_{j_{k+l}}|u_{i_{k+l}}\rangle_A \nonumber \\
&\kern-30pt \prod_{n=k+l+1}^{N} \delta_{j_{n} i_{n}} +\ldots+ (-1)^{N-k} \langle u_{j_{k+1}}|u_{i_{k+1}}\rangle_A \ldots \langle u_{j_{N}}|u_{i_{N}}\rangle_A \Bigg] \nonumber \\
&\kern-40pt = \frac{{N \choose k}}{N!} \sum_{i\rm{'s}, j\rm{'s}} \varepsilon_{i_1 \ldots i_N} \varepsilon_{j_1 \ldots j_N} |u_{i_1} \ldots u_{i_k}\rangle_A\ {}_A\langle u_{j_1} \ldots u_{j_k}| \times \nonumber \\
&\kern-30pt \sum_{l=0}^{N-k} (-1)^l {N-k \choose l} \langle u_{j_{k+1}}|u_{i_{k+1}}\rangle_A \ldots \langle u_{j_{k+l}}|u_{i_{k+l}}\rangle_A \delta_{j_{k+l+1} i_{k+l+1}} \ldots \delta_{j_{N} i_{N}} \nonumber
\end{eqnarray}
In the end we get
\begin{eqnarray}\label{eq:rho-1}
\tilde{\rho}_{k,N-k} &=& \frac{1}{k!} \sum_{l=0}^{N-k} \frac{(-1)^l}{l!} \sum_{i\rm{'s}, j\rm{'s}} \delta_{i_1 \ldots i_{k+l}}^{j_1 \ldots j_{k+l}} |u_{i_1} \ldots u_{i_k}\rangle_A\ {}_A\langle u_{j_1} \ldots u_{j_k}| \times \nonumber \\
& & \langle u_{j_{k+1}}|u_{i_{k+1}}\rangle_A \ldots \langle u_{j_{k+l}}|u_{i_{k+l}}\rangle_A
\end{eqnarray}
where we have again used (\ref{eq:levicivita-kronecker}). The final expression is exactly same as (\ref{eq:rho-2}). This completes the proof $\tilde{\rho}_{k,N-k} = \rho^{(N)}_{A,k}$.

\section{Appendix B. Multiple matrices \label{app:multi}}

Let us consider the matrix model described by \eq{three-one}. The model is
supersymmetric and has bosonic and fermionic matrix variables:
$X^I_{ij}, \chi_{ij}$. We will first ignore the fermions in the
following discussions (i.e. consider the bosonic model) and
briefly discuss them later in the section.
In the $A_t=0$ gauge, the theory has a residual symmetry under the
time-independent SU(N) transformation. This is ensured by the Gauss
law condition 
\be
\sum_I [X^I, P_I]\Psi[X] =0, 
\label{mone}
\ee
on the wavefunctions. Eq. \eq{mone} is equivalent to the singlet condition
\footnote{A similar condition also applies to the fermions $\chi_{ij}$.
  \label{ftnt-gauss-fermions}}
\be
  \Psi[X^I]= \Psi[U X^I U^\dagger].
\label{mtwo}
\ee
As a consequence of the $SU(N)$ invariance, we can make one of the matrices, say $X^1$, diagonal:
\[
X^1= D= {\rm diag}[\lambda_1, ..., \lambda_N].
\]
To do this, we
write $X^1$ in the form $X^1 = V D V^\dagger$, and make a change of
variables $X^1 \to (V, D)$, $X^I \to {\tilde X}^I= V^\dagger X^I V$,
$I=2,...,9$.  The $SU(N)$ amounts to demanding that the wavefunctions
are independent of $V$. The Jacobian of this change of variables is
the square of the Vandermonde determinant
\[
\Delta(\lambda) = \prod_{1\le i<j\le N}(\lambda_i - \lambda_j).
\]
In other words,
\be
\prod_{I=1,...,9}[dX^I] = \Delta^2(\lambda) \prod_{i=1,...,N}d\lambda_i
\prod_{I=2,...9} [d{\tilde X}^I] [dV]
\label{DxI}
\ee
The scalar product between two wavefunctions are given by
\bea
& \kern-50pt\int \kern-10pt \prod_{I=1,...,9}\kern-5pt [dX^I]\ \Psi^*[X^I] \Phi[X^I]
={\rm Vol}\left(SU(N)\right) \int\kern-10pt
\prod_{i=1,...,N} \kern-10pt d\lambda_i \ \Delta^2(\lambda)
\prod_{I=2,...9}\kern-5pt [d{\tilde X}^I]\ \Psi^*[D,{\tilde X}^I] \Phi[D,{\tilde X}^I]
\nonumber\\
& = \int\ [d\mu]  \tilde\Psi^*[D,{X}^I]\ \tilde \Phi[D,{X}^I]
\nonumber\\
& [d\mu]=  \prod_{i=1,...,N}\kern-5pt d\lambda_i \
\prod_{I=2,...9}\kern-5pt [d{X}^I]
\label{measure}
\eea
where
\be
\tilde\Psi[D,X^I]= C \Delta(\lambda) \Psi[D,X^I],\;I=2,...,9.
\label{tilde-psi}
\ee
The constant $C= \sqrt{{\rm Vol}\left(SU(N)\right)}$. In the first step, we have used the 
measure \eq{DxI} and the singlet condition \eq{mtwo} on the wavefunctions, so that the SU(N) transformation matrix $V$ simply comes out of the integral, yielding a volume factor. In the second step we have absorbed the Vandermonde determinant in each wavefunction, to have a simpler flat measure $[d\mu]$.

\gap2

\noindent{\it Residual symmetry: Weyl transformation}

\gap2

Even after fixing $X^1$ diagonal, there is a residual transformation, represented by the Weyl group $S(N) \subset SU(N)$, which permutes the eigenvalues
\be
(\l_1,\l_2,...,\l_N) \mapsto (\l_{\sigma(1)},\l_{\sigma(2)},...,\l_{\sigma(N)}),\, \sigma \in S(N).
\label{permute-l}
\ee
Under the transformation $\sigma$, we also have
\be
X^I \mapsto
\sigma(X^I),\; X^I_{ij}= X^I_{\sigma(i)\sigma(j)}, \, I=2,...,9.
\label{permute-x}
\ee

For a simple example, for $N=2$, we have
\[
X^1= \pmatrix{\l_1 & 0 \cr 0 &\l_2},\;
X^2 =\pmatrix{x^2_{11} & x^2_{12}\cr x^2_{21} & x^2_{22}},\; ...
\]
The ... represent $X^3$ onwards which have a similar expression.
The Weyl group is $S(2)$ which is generated by the single $SU(2)$
transformation matrix
\[
S= \pmatrix{0 & i\cr i & 0}.
\]
which represents the permutation $\sigma: (1,2)\mapsto (2,1)$.
It is easy to compute $\sigma(X^I):= S X^I S^\dagger$ for all $I=1,...,9$.
We find
\bea
&\sigma(X^1)= \sigma\left[\pmatrix{\l_1 & 0 \cr 0 &\l_2}
  \right]=\pmatrix{\l_2 & 0 \cr 0 &\l_1},\nonumber\\
& \sigma(X^2) =\sigma\left[\pmatrix{x^2_{11} & x^2_{12}\cr x^2_{21} & x^2_{22}}
  \right]= \pmatrix{x^2_{22} & x^2_{21}\cr x^2_{12} & x^2_{11}},\; ...
\label{weyl}
\eea
which confirms \eq{permute-l} and \eq{permute-x}.

We must ensure that the wavefunctions are also invariant under these
residual (Weyl) transformations, as required by \eq{mtwo}.  In the
$N=2$ case, this condition, in the diagonal $X^1$ gauge, becomes
\[
\Psi[\l_1, \l_2; x^2_{11}, x^2_{12}, x^2_{21}, x^2_{22}; ...]
= \Psi[\l_2, \l_1; x^2_{22}, x^2_{21}, x^2_{12}, x^2_{11}; ...]
\]
In the \eq{tilde-psi} basis, we will have
\be
\tilde\Psi[\l_1, \l_2; x^2_{11}, x^2_{12}, x^2_{21}, x^2_{22}; ...]
=- \tilde\Psi[\l_2, \l_1; x^2_{22}, x^2_{21}, x^2_{12}, x^2_{11}; ...]
\label{permute-tilde-psi-n-2}
\ee
where the $-$ sign appears because of the Vandermonde determinant
$\Delta= (\l_1 - \l_2)$ which picks up a minus sign under the permutation
$(1,2)\mapsto (2,1)$.

For more general $N$, the above equation \eq{permute-tilde-psi-n-2}
becomes, for all $\sigma \in S(N)$
\be
\tilde\Psi[\l_i; x^2_{ij}, x^3_{ij},...,x^9_{ij}]
= {\rm sign}(\sigma) \tilde\Psi[\l_{\sigma(i)}; x^2_{\sigma(i)\sigma(j)},
x^3_{\sigma(i)\sigma(j)},...,x^9_{\sigma(i)\sigma(j)}]
\label{permute-tilde-psi}
\ee

For the case of the single matrix, the above equation simply becomes
the statement that the wavefunction $\tilde\psi$ represents $N$ fermions.
This was the case discussed in Appendix A.

\gap2

\subsection{{\bf Target space EE for multiple matrices}\label{target-EE-matrices}}

\gap2

We will now discuss how to define target space EE for the model of D0
branes \eq{three-one}. The variables of the theory are the matrices
$\{X^I_{ij}\}/S(N)$ where the quotient represent dividing by the Weyl
transformations. In the diagonal $X^1$ gauge, the wavefunctions, satisfying \eq{permute-tilde-psi},
can be written as
\bea &\kern-60pt\tilde\Psi=\tilde\Psi_0(\l_1,...,\l_N;
X^2_{11},X^2_{12},...,X^2_{NN};...)  + {\rm Weyl~
  transforms}\nonumber\\
&\kern-75pt=\tilde\Psi_0(\l_1,...,\l_N;
X^2_{11},X^2_{12},...,X^2_{NN};...)  +\sum_{\sigma} {\rm sign}(\sigma)
\tilde\Psi_0[\l_{\sigma(i)}; x^2_{\sigma(i)\sigma(j)},
  x^3_{\sigma(i)\sigma(j)},...,x^9_{\sigma(i)\sigma(j)}]
\label{weyl-transform}
\eea
where the sum over $\sigma$ denotes all permutations of $S(N)$
(besides the identity). These are the equations \eq{three-seven} in
Section \ref{d0qm}. It is easy to see that the operators in the Hilbert space of such
wavefunctions are  given by \eq{three-eight}.

Let us imagine that we are interested in the target space region $A:
x^1 \ge d$.\footnote{\footnotesize Note that we are using the same
  notation $d$ as in the supergravity calculations. As indicated in
  the text (see discussions in Section \ref{d0qm}, a couple of
  paragraphs below \eq{three-eight}) in general these two quantities
  need not be identical; however, the difference between the two can
  be neglected when $d$ is sufficiently large in an appropriate
  sense.} What is the target space entanglement entropy corresponding
to such a region? In particular, how do we generalize the concepts
of Appendix A (Section \ref{appendix}) to a theory of {\it matrices}?

\gap2

\noindent\underbar{\it Classical moduli space}

\gap2

Note that there is no easy way to associate configurations of $N
\times N$ matrices to regions of target space. {\it A priori} the
simple $SU(N)$-invariant objects are traces of these matrices and
their products. In the diagonal $X^1$ gauge, the eigenvalues $\l_i$,
are also invariant objects, upto permutation, which can be mapped to
points on the $x^1$ axis. How does one construct a $d$-dimensional
region $A$ defined by the codimension one hypersurface $x^1>d$?

To get an idea, let us turn to the classical moduli space of the D0
brane matrix model \eq{three-one}, which corresponds to solutions of
the equation $[X^I, X^J]=0$. By analogy with higher dimensional gauge
theories, we will call this moduli space the `Coulomb branch'. In the
diagonal gauge for $X^1= {\rm diag}[\l_1, \l_2,..., \l_N]$, this
implies $X^I= {\rm diag}[X^I_{11}, X^I_{22}, ...,
  X^I_{NN}]$;$I=2,...,9]$. The solutions ${\bf x}_i= (\l_i, X^I_{ii})$
represent the coordinates of the $N$ D0 branes, $i=1,2,...,N$. Because
of the Weyl invariance, the classical moduli space of D0 branes is
\be
{\cal M}= \frac{{\bf R}^D}{S(N)}, \; D=9
  \label{moduli}
\ee
which is the same as that of $N$ indistinguishable particles in
${\bf R}^D$. Thus one can define a `classical sector' of
configurations where $k$ out of the $N$ identical particles are in the
$D$-dimensional region $A \subset {\bf R}^D$ (and the remaining
$N-k$ in $\bA$) (see Figure \ref{fig-coulomb-branch}). How does one
proceed to the quantum theory?
\footnote{Note that
  ordinarily in a 0+1 dimensional theory, the classical moduli space
  is not expected to survive under quantization since there is no
  spontaneous symmetry breaking. However the situation with D0 branes
  is somewhat subtle, especially because of supersymmetry; for an
  early discussion, see \cite{Taylor:1997dy} and references therein.}

Note that {\it \underbar{a} quantization} of the classical
configuration space \eq{moduli} was presented in Appendix A which
discussed the case of $N$ fermions in $d$-dimensions (see in
particular \eq{h-n},\eq{psi-p-q-d}). Following along the same lines,
we could try defining different sectors of the Hilbert space of wavefunctions
\eq{weyl-transform} by projecting onto mutually exclusive subspaces in
which $k$ number of ${\bf x}_i$'s in the region $A$ (the remaining
$N-k$ being in $\bA$), $k=0,1,...,N$, yielding a similar decomposition
as in \eq{h-n}:
\be
\mH = \oplus_{k=0}^N \mH_{k,N-k}
\label{sectors-matrices}
\ee

In the above, we have {\it defined} ${\bf x}_i$ = $(\l_i, X^I_{ii})$.
As explained in Figure \ref{fig-coulomb-branch}, these variables
define quantum fluctuations of the coordinates (which are equivalently
described in terms of open strings).

This is not yet a full specification of the quantum theory since we
have not said what to do with the extra, {\it off-diagonal}, variables
$X^I_{ij}$, which were not present in the $N$-particle problem. As
explained in Figure \ref{fig-coulomb-branch} these represent open
strings connecting different branes. Out of these, there are open
strings which connect the $k$ D0 branes which are all in region $A$
(these are not present in the Figure since $k=1$ there).  We should
definitely include them among our observables (i.e. include them in
our operator algebra); similarly the open strings which connect
different the $N-k$ D0 branes should be excluded from the operator
algebra. The issue is what to do with open strings that straddle
between region $A$ and $\bA$. Depending on the choice we make,
we arrive at two proposals (see Figure \ref{fig-coulomb-branch}):\\

\begin{itemize}

\item[Proposal 1:] We exclude the variables $X^I_{ij}$ straddling between region $A$ and
$\bA$ from the operator algebra. This leads to \eq{three-nine}
of Section \ref{d0qm}.
  
\item[Proposal 2:] We take the variables $X^I_{ij}$ straddling between region $A$ and
$\bA$ as part of the operator algebra. This leads to \eq{three-12}
  of Section \ref{d0qm}.

\end{itemize}

\end{document}